\documentclass{aa}  

\usepackage{graphicx}
\usepackage{txfonts}
\usepackage{natbib}  

\bibpunct{(}{)}{;}{a}{}{,} 
\usepackage{hyperref}
\newcommand{\lya}{Ly~$\alpha$}
\newcommand{\ha}{H\,$\alpha$}

\newcommand{\HI}{H\,$\textsc{i}$}
\newcommand{\OII}{[O\,$\textsc{ii}$]}

\begin{document}

   \title{Trident: A three-pronged galaxy survey}

   \subtitle{I. Lyman alpha emitting galaxies at z$\sim$2 in GOODS North \thanks{Based on observations made with the Nordic Optical Telescope, operated by the Nordic Optical Telescope Scientific Association at the Observatorio del Roque de los Muchachos, La Palma, Spain, of the Instituto de Astrofisica de Canarias.} }

   \author{A. Sandberg
          \inst{1}
          \and
          L. Guaita\inst{1}
          \and
          G. {\"O}stlin\inst{1}
          \and
          M. Hayes\inst{1}
          \and
          F. Kiaeerad\inst{2,3}
          }

   \institute{The Oskar Klein Centre, Department of Astronomy, Stockholm University, AlbaNova, 106 91 Stockholm, Sweden.\\
   \email{sandberg@astro.su.se}
         \and
             Aarhus University School of Engineering, Finlandsgade 22, 8200 Aarhus
     \and
     Nordic Optical Telescope, Rambla Jos\'{e} Ana Fern\'{a}ndez P\'{e}rez 7, 38711,
Bre\~{n}a Baja, Spain
             }

   \date{}

 
  \abstract
   {Lyman alpha (\lya ) emitting galaxies (LAEs) are used to probe the distant universe and are therefore important for galaxy evolution studies and for providing clues to the nature of the epoch of reionization. However, the exact circumstances under which \lya\ escapes a galaxy are still not fully understood.}
   {The Trident project is designed to simultaneously examine \lya , \ha , and Lyman Continuum emission from galaxies at redshift z$\sim$2, thus linking together these three aspects of ionising radiation in galaxies. In this paper, we outline the strategy of this project and examine the properties of LAEs in the GOODS North field. }
   {We performed a narrowband LAE survey in GOODS North using existing and two custom made filters at the Nordic Optical Telescope with MOSCA. We use complementary broad band archival data in the field to make a careful candidate selection and perform optical to near-IR SED fitting. We also estimate far-infrared luminosities by matching our candidates to detections in Spitzer/MIPS 24$\mu$m and Herschel/PACS catalogs.}
   {We find a total of 25 LAE candidates, probing mainly the bright end of the LAE luminosity function with $L_{Ly \alpha} \sim 1-15 \times 10^{42}$ erg s$^{-1}$. They display a range of masses of $\sim 0.5 - 50 \times 10^9 M_{\odot}$, and average ages from a few tens of Myr to 1 Gyr when assuming a constant star formation history. The majority of our candidates also show signs of recent elevated star formation. Three candidates have counterparts in the GOODS-Herschel far-IR catalogue, with luminosities consistent with ultra-luminous infrared galaxies (ULIRGs). }
   {The wide range of parameters derived from our SED fitting, as well as part of our sample being detected as ULIRGs, seems to indicate that at these \lya\ luminosities, LAEs do not necessarily have to be young dwarfs, and that a lack of dust is not required for \lya\ to escape.}

   \titlerunning{Trident I: LAEs in GOODS North}
   \maketitle
%

\section{Introduction}

Few emission lines have been studied as extensively in the distant universe as 1216 \AA\ (\lya). First suggested by \citet{partridge-peebles1967} as a means of finding high redshift galaxies, it became a signature to look for in galaxy surveys for decades. It was however not until deep dedicated surveys were performed that any significant quantities of these \lya\ emitting (LAE) galaxies were found \citep{cowie-hu1998}. Ever since, \lya\ has been successfully used to find galaxies across redshifts $z \sim 2-7$ \citep[e.g.][]{rhoads-2000, kudritzki-2000, malhotra-rhoads2002, ouchi-2003, ouchi-2005, gawiser-2006, ajiki-2006, gronwall-2007, pirzkal-2007, finkelstein-2008, nilsson-2009, hayes-2010a, yuma-2010, ouchi-2010,guaita-2011,adams-2011,blanc-2011,shibuya-2012,vargas-2014},
even probing the end of the epoch of reionization \citep[e.g.][]{malhotra-rhoads2004,dijkstra-2007,ono-2010,jensen-2013,matthee-2014}.   

   It quickly became apparent in the earliest studies that \lya\ is nowhere near as bright and ubiquitous in high redshift galaxies as was first imagined (see for example \citet{pritchet1994} for a review of non-successful \lya\ surveys). Recombination theory tells us that roughly one third of all of the ionizing flux from hot stars should be reprocessed into the \lya\ emission line as the atomic hydrogen gas recombines \citep{osterbrock1962}, but something is keeping \lya\ from escaping most galaxies.
   
   It was natural to assume that interstellar dust would play an important role in absorbing \lya\ photons, as dust absorbs rest-frame ultraviolet light very effectively. Indeed, early studies showed a tentative anti-correlation between dust content and \lya\ emission \citep{meier-terlevich1981}. However, it was soon shown that this could not be the only factor governing \lya\ escape, as there are examples of strongly emitting \lya\ galaxies rich in dust, and nearly dust-free galaxies where \lya\ is completely absorbed \citep{kunth-1994,giavalisco-1996}. \citet{kunth-1998} showed the importance of the interstellar medium (ISM) kinematics, as the \lya\ photons resonantly scatter on neutral \HI\ gas. \lya\ radiation at the rest wavelength is optically thick already at column densities $\mathrm{N_{H\,\textsc{i}}} \sim 10^{14}$ cm$^{-2}$, so even a tiny amount of neutral hydrogen will affect the emergent line significantly. With multiple scatterings, the path length is greatly increased, and even a small amount of dust can in principle absorb a large amount of photons. However, if the neutral \HI\ gas is in rapid motion compared to the \lya\ sources, for example due to supernova feedback or strong stellar winds, \lya\ can be shifted out of resonance and escape more easily. With a large range of observational studies, combined with advanced simulations of \lya\ radiative transfer through galaxies \citep[e.g.][]{verhamme-2006,verhamme-2012,laursen-2013,duval-2014}, we are only now starting to build a coherent picture of how \lya\ is produced and escapes from galaxies.
   
   There are still, however, many questions that remain unanswered. Two galaxies with essentially the same properties in terms of mass, star formation rate, age, gas-mass ratio, dust content etc., can show completely different behavior when it comes to \lya\ \citep[e.g.][]{hayes-2014}. We are still not certain of which parameters are dominant for governing \lya\ escape. As the number of LAE surveys is rapidly increasing, and the volumes and numbers probed can now be measured in whole degrees on the sky and amount to thousands of galaxies, it is more important than ever that we truly understand the mechanisms governing these galaxies. 
   
   Another outstanding question in the study of high redshift galaxies is the manner of escape of photons with an energy above 13.6 eV ($<$912 \AA), capable of ionizing neutral hydrogen. We refer to this as Lyman continuum (Ly C) radiation in this paper. As the universe was gradually reionized somewhere at $z \gtrsim 6$ \citep{fan-2006}, the natural ingredient that is assumed responsible are the first stars and galaxies. However, in order to reionize all of space, the amount of ionizing radiation escaping out of these galaxies must have been very large. Yet, when we try to measure this quantity in strongly star-forming galaxies ("starbursts"), we see relatively modest escape fractions \citep[see e.g.][and references therein]{vanzella-2012}. The Ly C escape fraction remains difficult to measure directly, as the signal from nearby sources is absorbed by Galactic \HI\ gas and requires space-based ultraviolet observatories with high sensitivity even at redshifts up to $z \lesssim 3$.  
   
   With these unresolved questions in mind, we started the project that we call Trident.  In this first paper, we present the general outline of the project and our first results, focusing on \lya\ emitting galaxies in GOODS-N. The \ha\ to \lya\ properties and the Ly C escape will be the subject of future papers. In this paper, we focus on the properties of the \lya\ emission, estimating stellar properties from spectral energy districution (SED) analysis. We also identify three individual candidates in far-infrared survey data and discuss the implications on their dust content.
   
   In section~\ref{sec:trident} we outline the general idea of Trident and how it is structured.
   In section~\ref{sec:data} we describe the data we have collected and how it was reduced. Section~\ref{sec:candsel} describes how we selected our candidates from the data. In section~\ref{sec:FIR} we measure far infrared properties of our candidates found in the GOODS-Herschel catalog. In section~\ref{sec:SED} we explain how we combined our data with archival survey data of GOODS-N, and how this was then used for the SED fitting routine. In section~\ref{sec:conclusions} we summarize and discuss the implications of these results. 
   
   Throughout the paper, we assume H$_0 = 70~\mathrm{km~s^{-1} Mpc^{-1}}$, $\Omega_M = 0.3$, $\Omega_\Lambda = 0.7$

   \section{The Trident survey}\label{sec:trident}
   
   We chose the name Trident because we are trying to find three types of emission related to atomic hydrogen from the same galaxies in one combined survey. The idea for a combined \lya\ and \ha\ study builds largely on the Double-Blind Deep project \citep{hayes-2010a,hayes-2010b}, in which these emission lines were successfully studied in galaxies in the GOODS-South field, using the FORS1 and HAWK-I instruments at ESO VLT. Being a relatively deep study (reaching line fluxes at 5$\sigma$ of 6.8 and $7.8\times10^{-18} \mathrm{erg s^{-1} cm^{-2}}$ in \ha\ and \lya , respectively), it probed the faint end of the \lya\ luminosity function (LF) at $z \sim 2$ to an unprecedented accuracy. A similar approach was used by \citet{nakajima-2012} who studied \lya , \OII\ $\lambda 3727$ and \ha\ galaxies at $z \sim 2.2$ in the SXDS field.
   
   Drawing on our experience gained from Double-Blind Deep, we started our current project. One of the drawbacks of Double-Blind Deep was that it was designed to find predominantly faint emitters. However, constraining the bright end of the \lya\ LF requires a larger survey volume, as bright emitters are rare, and this is one of the goals of the current project. In this paper, we present our first results from a shallow survey over a large volume, specifically targeting the brightest \lya\ emitters. We are thus not expecting to find large numbers of candidates, but rather we are interested in their \ha\ to \lya\ properties and use ancillary information in well studied fields to connect these LAE candidates to other galaxy populations.
   
   We target well studied extragalactic fields to maximize the available multi-wavelength data, including observations with the Hubble Space Telescope (HST) in the ultraviolet, in order to add the third prong to our trident. In addition to the GOODS-North observations presented here, we are currently observing with NOT/MOSCA two Frontier Fields and Abell\,1689 -- lensing clusters with existing deep HST UV observations -- using several new optical filters with deeper exposures than those presented here ($\sim$ 10 hours per narrowband pointing). The total volumes probed per cluster will be similar to the central pointing of GOODS-North presented here, but reaching fainter emitters, also possibly magnified by the foreground clusters. We have started gathering \ha\ observations of Abell\,1689 in the near-infrared, and we plan to observe our two Frontier Fields with CFHT/WIRCam in the near future. Once these observations are completed, they will be the subject of future Trident project papers. 
   
   Observing fields with HST UV coverage allows us to investigate the leakage of Ly C photons from the galaxies we detect in the survey. These ionizing photons are intimately linked to the \lya\ emission, as they are required for \lya\ to form. The high absorption cross-section in \HI\ for both types of emission may cause them to escape together, probably through ionized cones or channels with low column densities and/or covering fractions, as models of porous multi-phase media suggest \citep[e.g.][]{clarke-oey-2002,neufeld-1991,laursen-2013,verhamme-2014}. Observations and models of intermediate redshift galaxies both support this picture, as LAE galaxies appear to emit more Ly C photons than what Lyman break galaxies (LBGs) do \citep{iwata-2009,nestor-2013,mostardi-2013}. Models also suggest that escape fractions of Ly C are higher in lower-mass haloes \citep{yajima-2011,mitra-2013}. Note that all LAEs must show a Lyman break, and that selection effects will make an LAE survey find relatively more low-mass galaxies on average for a given exposure time. 
   
   Adding to this the observed steep UV LFs of faint galaxies at $z \gtrsim 2$ \citep{alavi-2014}, it therefore seems possible that low-mass \lya\ emitting galaxies did have large enough Ly C escape fractions to ionize the universe \citep[see e.g.][]{bouwens-2009,bouwens-2012,robertson-2013}, and recent observations of the faint end of the \lya\ and UV LFs at $z\sim6-8$ are well in line with this picture \citep{dressler-2014,atek-2015}. However, due to the rapidly increasing opacity of the IGM beyond $z \gtrsim 4$ \citep{inoue-iwata-2008,inoue-2014}, this Ly C escape can never be directly observed. Linking the \ha , \lya\ and Ly C emission together at $z \sim 2$ is therefore one of very few ways of quantitatively studying Ly C escape in galaxies.  
   
\section{Observations and data reduction}\label{sec:data}

\begin{figure}
   \centering
 \includegraphics[width=\hsize]{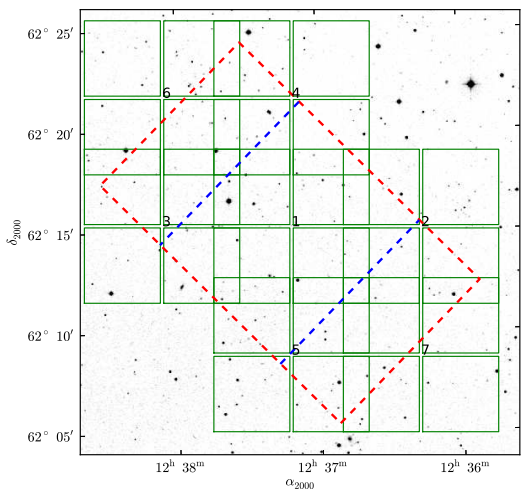}
    \caption{Fiducial map of pointings for MOSCA with filter NB401\_6. Numbers correspond to pointings as described in Table~\ref{tab:obslog}. Dashed lines outline the GOODS ACS (red) and the CANDELS-Deep (blue) footprints.}
              \label{fig:MOSCA_FOV}%
    \end{figure}

\subsection{Custom filters}

   The general idea of finding a galaxy in a narrowband emission line survey is to do photometry of a large piece of the sky, using a narrowband filter where the wavelength is centered on a redshifted emission line. When comparing the brightness of an object in this narrowband filter to a nearby broadband filter, anything with a strong emission feature in the narrowband filter will show a narrowband color excess \citep[see e.g.][for recent descriptions of the narrowband technique and how it can be used]{guaita-2010,nakajima-2012,sobral-2013}. This therefore gives a signature that can be used to find candidates for follow-up studies. 
   
   One can of course combine narrowband filters at different wavelengths in order to see different emission lines. In our case, we want to look for galaxies not only in \lya\ but also in \ha , in a way that the redshift range is the same for both surveys. We therefore attempted to match near-IR narrowband filters with optical filters in terms of the redshifts of these emission lines. To this end, we had two custom filters made to exactly match existing near-IR filters. We have also used three other, pre-existing optical filters, which extends our \lya\ redshift range and thus increases the volume probed. We refer to the narrowband filters as e.g. NB392\_6, where 392 and 6 are the central wavelength and width, in nanometers. NB392\_6 and NB401\_6 were designed to match $\mathrm{H_2}$ and Brackett $\gamma$ near-IR filters. The properties of all the narrowband filters, and how they give an almost complete redshift coverage between $z=2.2$ and 2.46, are summarized in Table~\ref{tab:filters}. 
   
   Our \lya\ observations were performed at the 2.56 meter Nordic Optical Telescope (NOT) on La Palma, Spain, using the MOSCA instrument. MOSCA is a mosaic camera that is mounted at the Cassegrain focus, and it consists of four CCDs laid out to form a roughly 7.5 arcmin wide field of view. MOSCA's CCDs are especially sensitive in the blue (with a quantum efficiency near 97\% at 430 nm), and with few internal surfaces it is an ideal instrument for our purposes. 
   
   We observed GOODS-North, using a central pointing of $\mathrm{12^{h} 36^{m} 54^{s}, +62^{\circ} 14\arcmin 50\arcsec}$ (J2000). For NB401\_6, we observed a grid of positions shown in Fig.~\ref{fig:MOSCA_FOV}, covering a surface area of $\sim$ 400 square arcminutes. Table~\ref{tab:obslog} lists the exposure times and seeing conditions for each pointing. The other narrowband filters where centered on the same coordinates but rotated 45 degrees to better match with the HST/ACS data footprint. The corresponding volume that we cover in the NB401\_6 filter is roughly 83 000 comoving Mpc$^3$, with an additional $\sim 60 000$ Mpc$^3$ in supplementing filters. This is comparable to e.g. the \citet{guaita-2010} MUSYC study, which covers $\sim$ 135 000 Mpc$^3$ of LAEs at $z\sim2.07$, but goes considerably deeper.
    
    The \ha\ data for GOODS-North cover circa 22 arcmin$^2$ (roughly the size of the view in Figure~\ref{fig:MOSCA_FOV}) in a narrowband filter matching the NB401\_6 filter in redshift space. The details of these observations are described in a forthcoming paper. 
   
\begin{table}
\caption{Narrowband filter properties}             
\label{tab:filters}      
\centering                          
\begin{tabular}{c c c c c c}        
\hline\hline                 
Filter & ID & $\lambda_c$ (\AA) & $\lambda_w$ (\AA) & $z$ range & ZP \\    
\hline                        
   NB392\_6 & \#134 & 3917 & 63  & $2.20-2.25$ & 22.70 \\
   NB401\_6 & \#133 & 4013 & 64  & $2.27-2.33$ & 22.72 \\
   NB406\_6 & \#121 & 4058 & 58  & $2.31-2.36$ & 22.52 \\
   NB410\_6 & \#131 & 4102 & 60  & $2.35-2.40$ & 22.37 \\
   NB412\_18 & \#115 & 4120 & 180 & $2.32-2.46$ & 23.66\\ 
\hline                                   
\end{tabular}
\tablefoot{Nordic Optical Telescope ID number, central wavelength, width and corresponding approximate redshift range of \lya\ for the narrowband filters used in this study. The final column lists the average zeropoint for each filter, in AB magnitudes, acquired with a $\sim$ 6 arcsec aperture. The uncertainty in all zeropoints is roughly 0.05 mag.}
\end{table}

   \begin{figure}
   \centering
   \includegraphics[width=\hsize]{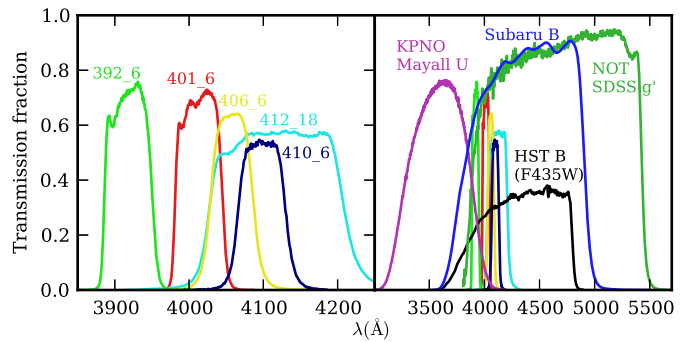}
      \caption{Transmission curves for the narrowband NOT filters used in this work (left panel) and how they relate to broad U, B and g' filters (right panel). Note that the HST F435W filter is scaled to the entire instrument throughput.
              }
         \label{fig:filtercurves}
   \end{figure}
   
\subsection{Data reduction}   
   
   The NOT/MOSCA data discussed in this paper were taken in March and May 2011 and February and May 2012. Table~\ref{tab:obslog} summarizes these observations.
   
   \begin{table}
\caption{Observing log}             
\label{tab:obslog}      
\centering                          
\begin{tabular}{c c c c c}        
\hline\hline                 
Date & Filter & Pointing & Exp. time & Seeing \\    
\hline                        
   2011-03-01 & NB406\_6 & 1* & 2.5h & $0\farcs9 - 1\farcs1$  \\  
   2011-03-01 & NB410\_6 & 1* & 1.5h & $1\farcs1 - 1\farcs2$  \\ 
   2011-03-02 & NB412\_18 & 1* & 1.5h & $0\farcs8 - 1\farcs0$  \\
   2011-03-02 & NB406\_6 & 1* & 1.5h & $0\farcs8 - 0\farcs9$  \\
   2011-03-02 & NB410\_6 & 1* & 1.5h & $0\farcs7 - 0\farcs8$  \\
   2011-05-02 & NB401\_6 & 1 & 2.5h & $0\farcs8 - 0\farcs9$  \\
   2011-05-03 & NB392\_6 & 1* & 3.5h & $0\farcs7 - 0\farcs8$  \\
   2011-05-04 & NB401\_6 & 1,2,3 & 2h,2h,0.5h & $0\farcs8 - 1\farcs0$  \\ 
   2011-05-05 & NB401\_6 & 2,3 & 0.5h,2h & $0\farcs8 - 1\farcs1$  \\ 
   2012-01-27 & NB401\_6 & 5 & 3h & $0\farcs9 - 1\farcs1$  \\
   2012-02-23 & NB401\_6 & 4 & 4h & $0\farcs9 - 1\farcs0$  \\
   2012-05-20 & NB401\_6 & 2 & 2h & $1\farcs3 - 1\farcs5$  \\
   2012-05-21 & NB401\_6 & 3,5 & 1.5h,1h & $0\farcs9 - 1\farcs5$  \\
   2012-05-22 & NB401\_6 & 6,7 & 2h,2h & $0\farcs7 - 0\farcs8$  \\
\hline                                   
\end{tabular}
\tablefoot{Log of the NOT/MOSCA Observations. The filters listed are only the narrowband filters used in observations; complementing broad band observations (usually in the SDSS g' band) where always performed. The "pointing" column corresponds to the numbers in Fig.~\ref{fig:MOSCA_FOV}. An asterisk is used to mark if the pointing was observed with a 45 degree field rotation.}
\end{table}
   
   Many of the basic reduction steps were developed from Tapio Pursimo’s notes\footnote{\url{http://www.not.iac.es/instruments/mosaic/reductions/reduction.notes}} on reducing MOSCA data using the IRAF\footnote{IRAF is distributed by the National Optical Astronomy Observatories, which are operated by the Association of Universities for Research in Astronomy, Inc., under cooperative agreement with the National Science Foundation.} MSCRED package. 
   
   After the basic bias, flat and dark reduction steps in MSCRED, the IRAF FIXPIX task was used to apply a bad pixel mask to the data. Next, the MSCSETWCS task from MSCRED was used to correct for individual rotation and distorsions in the four CCDs of MOSCA by changing the WCS projection. This step uses a plate solution which we constructed with the MSCTPEAK task.
    
   The sky level was then subtracted using Source Extractor \citep{bertin-arnouts1996}, and the images were combined to a single fits extension using the MSCIMAGE task. Cosmic ray cleaning was then applied using v0.4 of cosmics.py, a python interface to L.A.Cosmic\footnote{\url{http://obswww.unige.ch/\~tewes/cosmics\_dot\_py/}}. The MSCIMATCH task was used to match the intensity between the frames, taking into account differences in airmass and seeing conditions, etc. The WCS of each frame was matched against the USNO-A2.0 star catalogue with the MSCCMATCH task, which was set to shift the WCS to the correct position. 
   The science data were then stacked using the MSCSTACK task, which uses the WCS of the frames to align them before combining. 
    
   The final astrometric solution was obtained by running MSCCMATCH with a set of relatively bright HST b-band sources, in order to match our astrometry to that of \citet{giavalisco-2004} which is commonly used for sources in GOODS-North. We then used GREGISTER to register all of our ground-based data in all filters to a common set of physical image coordinates. This is a crucial step for doing the photometry with SExtractor in dual mode, where one image is used for detection and the other for photometry. 
   
   For our ground-based data, we use aperture corrected photometry where the correction is calculated from isolated bright sources in the field. We use aperture diameter sizes of 22 and 60 pixels, corresponding to radii of 2.3 and 6.3 arcsec. Detections were made using DETECT\_THRESH = 1.2 and DETECT\_MINAREA = 32 in SExtractor, meaning that anything with 32 contiguous pixels that are all 1.2 times brighter than the background is considered a detection. This can be compared to the seeing disk which for a full width at half maximum of 1 arcsec would correspond to an area of about 17 pixels. Defining the thresholds this way allows for low surface brightness detections with unknown morphologies.
   
   We estimate the depth of our NB401\_6 imaging to reach 24.3 (24.6) AB mag at the 90\% (50\%) completion level, for the areas with a total exposure time of $\sim$ 4.5 hours. This corresponds to a limiting \lya\ line flux (assuming a 20 \AA\ EW cut in the rest-frame) of $9.8~(7.3)\times10^{-17} \mathrm{erg~s^{-1}cm^{-2}}$. In the NE and SW outskirts of the NB401\_6 image, and in our other filters the effective exposure time is $\sim$ 2-3 hours, and the depth reached consequently 0.2-0.4 magnitudes shallower. 
   
\subsection{Standard stars}

   Our custom made filters require the use of spectrophotometric standard stars for the photometric calibration. We used white dwarf stars from \citet{oke-1990}, as these stars do not show a strong Balmer break around 4000 \AA. We corrected the \citet{oke-1990} absolute fluxes according to \citet[][usually a shift on the order of +0.05 AB magnitudes]{colina-bohlin1994}. We then convolved the standard spectra with our measured narrowband transmission curves and calculated the AB magnitude of each standard star in each of our filters. 
   
\begin{figure*}
   \centering
 \includegraphics{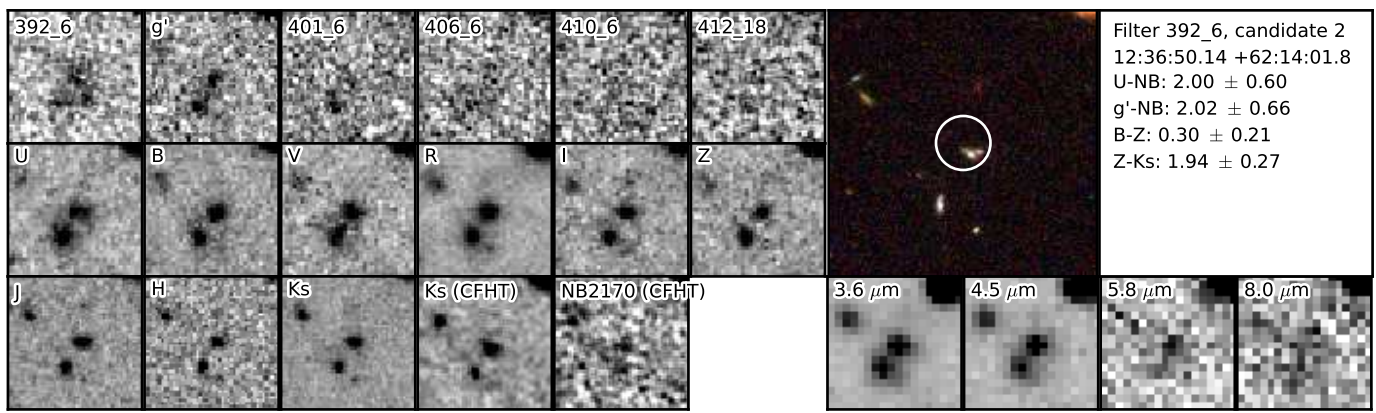}
   \caption{Postage stamps for an example candidate galaxy. The field of view is 10 $\times$ 10 arcsec. The large RGB picture shows an HST $bvi$ color composite made with the scale in each color channel related to the true total intensity, with an arcsinh intensity scaling relation. This coloring scheme is similar to that described in \citet{lupton-2004}. The white circle has a one arcsec radius - typically twice the size of the ground-based seeing disk. Postage stamps for all candidates are compiled in the Appendix.}
              \label{fig:cutouts}%
    \end{figure*}

   The CFHT/WIRCam/WC8305 narrowband image was reduced using the Simple-Wircam pipeline developed by Wei-Hao Wang \footnote{\url{http://www.asiaa.sinica.edu.tw/~whwang/idl/SIMPLE/}}. NB401\_6 is designed to match this filter, which was used for the matching \ha\ survey, in redshift space. The \ha\ results will be discussed in a forthcoming paper, but the postage stamps in Fig~\ref{fig:cutouts} show a cutout for reference purposes here.
    
\subsection{Archival and catalog data}\label{sec:catalog}

   A large set of multi-wavelength data already exists for various regions around GOODS-North. We have chosen to make use of ground-based $UBVRIz$ broadband imaging using the KPNO 4.0 meter telescope and Subaru 8.3 meter, described in \citet{capak-2004}. We use the R-band detection catalog, using auto photometry. We chose this photometric catalog over the HST/ACS catalog for several reasons. The covered area on the sky is larger, and includes the entire region that we examined with the NB401\_6 filter. The resolution is very similar to our data, as they are all ground-based, which makes the photometry easier to handle in a consistent manner. The KPNO U-band data is also important for our narrowband selection criteria.
    
   For our near-infrared data, we use the deep $K_s$ and Spitzer catalog from \citet{wang-2010}, who carefully matched Spitzer/IRAC data to each $K_s$ source. To fill in the $J$ and $H$ bands where available, we make use of the MOIRCS Deep Survey near-IR catalog \citep{kajisawa-2010}, and calculate a $J-K_s$ and $H-K_s$ color. The MOIRCS Survey covers a horizontal strip through a large portion of the field. 
   
   We do see a shift between the two $K_s$ catalogs of about 0.2$\pm$0.1 magnitudes. We investigated whether one of the two surveys would give a better optical-to-infrared color by measuring the average and median colors of all sources, but the end result is somewhere in between the two with quite a large uncertainty. We have therefore chosen to use the average value of the catalogs, and have added in quadrature an uncertainty of 0.2 magnitudes to all of our near-infrared and Spitzer data points. This reflects our uncertainty in the relative optical to near-infrared photometry.
    
\section{Candidate selection}\label{sec:candsel}

   \begin{figure}
   \centering
   \includegraphics[width=\hsize]{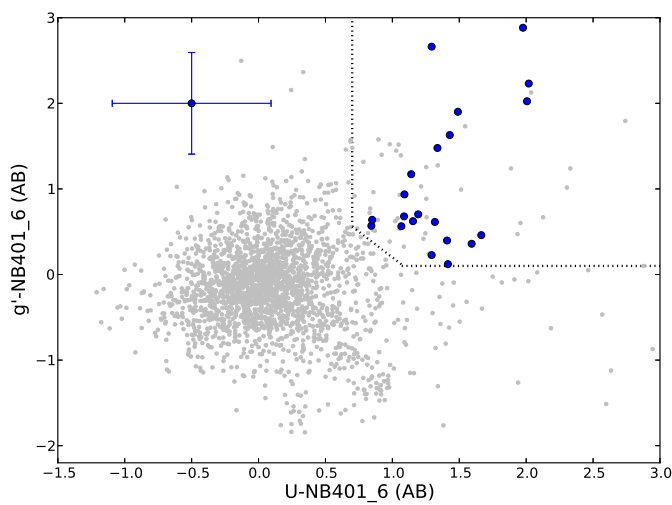}
      \caption{Narrowband selection for our NB401\_6 filter. Dotted lines show the cut-off. Photometry for all data is indicated by small gray points. Large blue points indicate candidates that meet both the excess and the BzK criteria. The blue data point on the top left shows the typical size of the uncertainty. 
              }
         \label{fig:Data_Unbg}
   \end{figure}
   
   \begin{figure}
   \centering
   \includegraphics[width=\hsize]{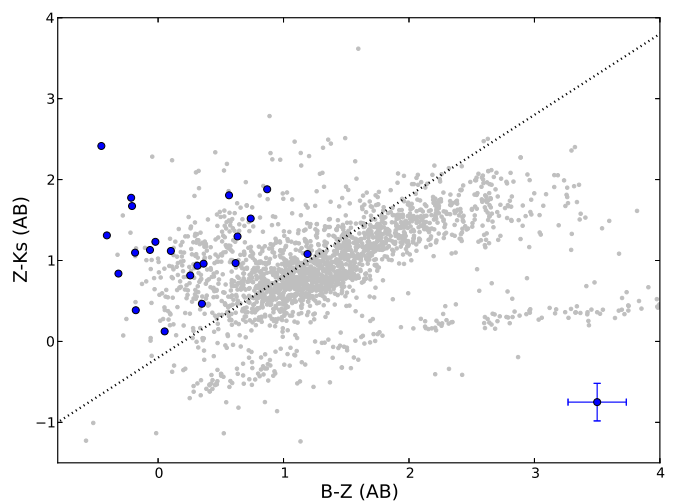}
      \caption{BzK selection for candidates in the NB401\_6 filter. The dotted line shows the cut-off, corresponding to the \citet{daddi-2004} BzK criterion.  Photometry for all data is indicated by small gray points. Large blue points indicate candidates that meet both the narrowband color selection and BzK criteria. The blue data point on the bottom right shows the typical size of the uncertainty.
              }
         \label{fig:Data_BzK}
   \end{figure}

There are several possible approaches to finding LAE candidates in a narrowband survey. A common way is to simply measure a narrowband to broadband color, where the broadband covers the nearby continuum. If a strong emission line falls within the narrow filter, this will produce a narrowband excess and possible candidates will therefore show a color signature. This method is relatively stable if the broadband is centered approximately on the narrowband, or if the continuum is expected to be flat throughout the probed wavelength range. 

   In this simplified case, the color signature in the narrowband filter can easily be estimated. The flux density in the narrowband filter can be described as 
   
   \begin{align}\label{ew_nb}
   f_{\nu,NB} = f_{\nu,NB}^{EL}+f_{\nu,NB}^{cont} = \frac{\int f_{\nu}^{EL}(\lambda)\frac{c}{\lambda^{2}}T_{NB}(\lambda)e^{-\tau_{IGM}(\lambda)}d\lambda}{\int\frac{c}{\lambda^{2}}T_{NB}(\lambda) d\lambda} &+ \\ \frac{\int f_{\nu}^{cont}(\lambda)\frac{c}{\lambda^{2}}T_{NB}(\lambda)e^{-\tau_{IGM}(\lambda)}d\lambda}{\int\frac{c}{\lambda^{2}}T_{NB}\left(\lambda\right)d\lambda}\notag
   \end{align}
   
   Where $f_{\nu,NB}^{EL}$ and $f_{\nu,NB}^{cont}$ are the flux density contributions from the emission line and continuum, $T_{NB}$ is the transmission curve of the filter, and the IGM transmissivity is described as a simple exponential with appropriate $\tau_{IGM}$ \citep[e.g.][]{madau-1995}. Such IGM absorption bluewards of \lya\ can be significant, especially at higher redshifts. The expression for the broadband filter will have an identical appearance. 
   
   The continuum can be described as a power law with slope $\alpha$, so that $f_{\nu}^{cont} = \frac{N}{c}\lambda^{\alpha}$, where $N$ is a normalization factor. If we assume the \lya\ line to be a delta function, and the IGM effects to be negligible, we can simplify \eqref{ew_nb} to
   
   \begin{align}\label{simpler_ew_nb}
   \qquad f_{\nu,NB}&=\frac{F_{EL}\overline{T_{NB}}}{\int\frac{c}{\lambda^{2}}T_{NB}(\lambda)d\lambda}+\frac{\int N \lambda^{\alpha -2}T_{NB}\left(\lambda\right)d\lambda}{\int\frac{c}{\lambda^{2}}T_{NB}\left(\lambda\right)d\lambda}
   \end{align}

   where $F_{EL}$ is the flux in the emission line and $\overline{T_{NB}}$ is the integral averaged transmissivity of the filter. Describing $F_{EL}$ in terms of the equivalent width, $F_{EL} = \mathrm{EW} \times N \lambda^{\alpha -2}$, we arrive (with \eqref{simpler_ew_nb}) at an expression for the flux in a given filter as a function of equivalent width and the continuum parameters $N$ and $\alpha$. The color for a given narrowband to broadband is then simply given by the usual magnitude definition,
   
   \begin{align}\label{fluxtomag}
   \qquad m_{NB} - m_{BB} &= -2.5 \log_{10}\left(\frac{f_{\nu,NB}}{f_{\nu,BB}}\right)
   \end{align}
   
   With a flat spectrum in $f_{\nu}$ ($\alpha = 0$) and an observed EW cut of $\sim$ 65 \AA\ (20 \AA\ in the rest-frame at $z \sim 2.3$), \eqref{fluxtomag} and the method described above would give us an excess with a color cut $g' - NB \approx 0.6$, or $U - NB \approx 0.7$.
   
   Note that the NB412\_18 filter is three times wider than the other filters. This means it is probing a three times wider redshift space but at the same time reaches a lower significance due to decreased contrast between continuum and line emission. We can only reliably resolve sources in this filter with an equivalent width in the observed frame similar to the width of the filter ($>$180 \AA).
   
   For a more realistic calculation, we need to take the complications mentioned above into account. Our narrowbands have wavelengths near the edges of our broadband filters, and moreover, the continuum on either side of \lya\ can not be expected to be flat. Absorption of \lya\ in \HI\ gas in the IGM will tend to attenuate the spectrum bluewards of \lya . A strong continuum slope can therefore create false candidates, or make real sources disappear, if a classic narrowband excess technique is used \citep[see e.g.][]{hayes-2006}. We have therefore chosen to combine two broadbands, the $U$ and $g'$  bands, on either side of \lya\ at $z \sim 2$. 
    
   In order to estimate our color selection criteria, we first created model spectra using our SED fitting code GalMC (see Section \ref{sec:SED}). We convolved these spectra with our measured filter transmission curves to calculate the expected brightness in each filter. 
   
   Using an EW(\lya ) cut of $\sim$ 20 \AA\ (rest-frame), we start by a color cut $U - NB > 0.7$, corresponding to the limit of no IGM absorption. This color also has a significance criterion, so that the photometric error is not larger than the color itself. To allow for reddening up to $E(B-V) = 0.5$, we set our limits at $g' - NB > 0.1$ and $g' - NB > -1.2 (U - NB) + 1.4$. Allowing for stronger reddening could introduce a group of interlopers at slightly different redshift, that may show a strong $U - NB$ excess due to the Lyman break, but have no actual \lya\ signal in the narrowband. Our limit is therefore chosen to balance these effects as well as possible. Many of our candidates have a large photometric uncertainty due to this being a shallow survey, and we therefore also chose this double-excess method to get a more reliable selection, although we will always have some interlopers in our candidate sample. The $-1.2$ slope was calculated from our GalMC models, and is set by the relative widths of the $U$ and $g'$ filters. 
   
   In order to deal with possible low redshift interlopers with \OII\ 3727 \AA\ emission, we combine these criteria with the standard $BzK \equiv (z - K) - (B - z) > -0.2$, designed to select star forming galaxies at $z > 1.4$. Figures~\ref{fig:Data_Unbg} and \ref{fig:Data_BzK} show these criteria applied to our NB401\_6 data. Using the BzK criterion requires a detection in these broadbands, and this criterion reduces our total number of LAE candidates from 42 to 30 sources. Given that the broadband data is considerably deeper than our narrowband data, and that we are looking specifically for bright emitters, we use this also as a criterion to remove spurious sources.
    
    We correlate our sample of candidates to the spectroscopic catalogues compiled by \citet{barger-2008} and \citet{adams-2011}. These catalogs only contain sources for which the redshift could be determined with several spectral features. We find five candidates to already have spectra consistent with the expected narrowband wavelengths (as indicated in Tables~\ref{tab:SEDtabA} - \ref{tab:SEDtabC}), but there are also spectroscopically discovered LAEs which are too faint to be detected in our sample. A thorough analysis of the \lya\ and \ha\ brightness of galaxies which are already spectroscopically confirmed will be included in a forthcoming paper. 
    
    We match our candidates to the Chandra Deep Field North X-ray survey \citep{alexander-2003} and deep VLA data \citep{morrison-2010} in order to remove AGN. Anything with an X-ray luminosity $>10^{42}$ erg/s or with a 20 cm detection $>10^{24}$ W/Hz is considered an AGN. We remove 5 candidates based on these criteria, leaving 25 LAE candidates for further analysis. Our AGN fraction of $\sim$ 17\% is broadly consistent with similar works (see Section~\ref{sec:previouswork}).
    
\section{Warm dust properties}\label{sec:FIR}

\begin{figure*}
   \centering
 \includegraphics[width=16.4cm]{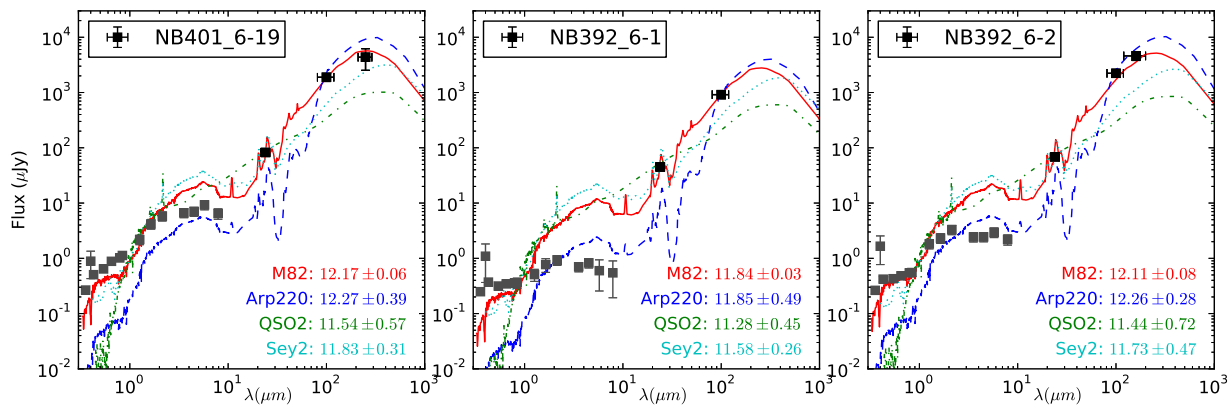}
    \caption{Fits to the far-infrared data points from the \citet{elbaz-2011} GOODS-Herschel catalog, using two templates of star forming galaxies (M 82 and Arp 220 in solid red and dashed blue, respectively) as well as two AGN templates (QSO2 and Seyfert 2 in dot-dashed green and dotted cyan). Note that only the infrared $>8 \mu$m catalog data were used in the fit, and the optical to near-infrared data are only shown here for comparison. The numbers in the bottom right of each panel show the integrated 8-1000 $\mu$m luminosities in solar units for each template.}
              \label{fig:FIRfitting}%
    \end{figure*}
    
    \begin{figure*}
   \centering
 \includegraphics[width=16.4cm]{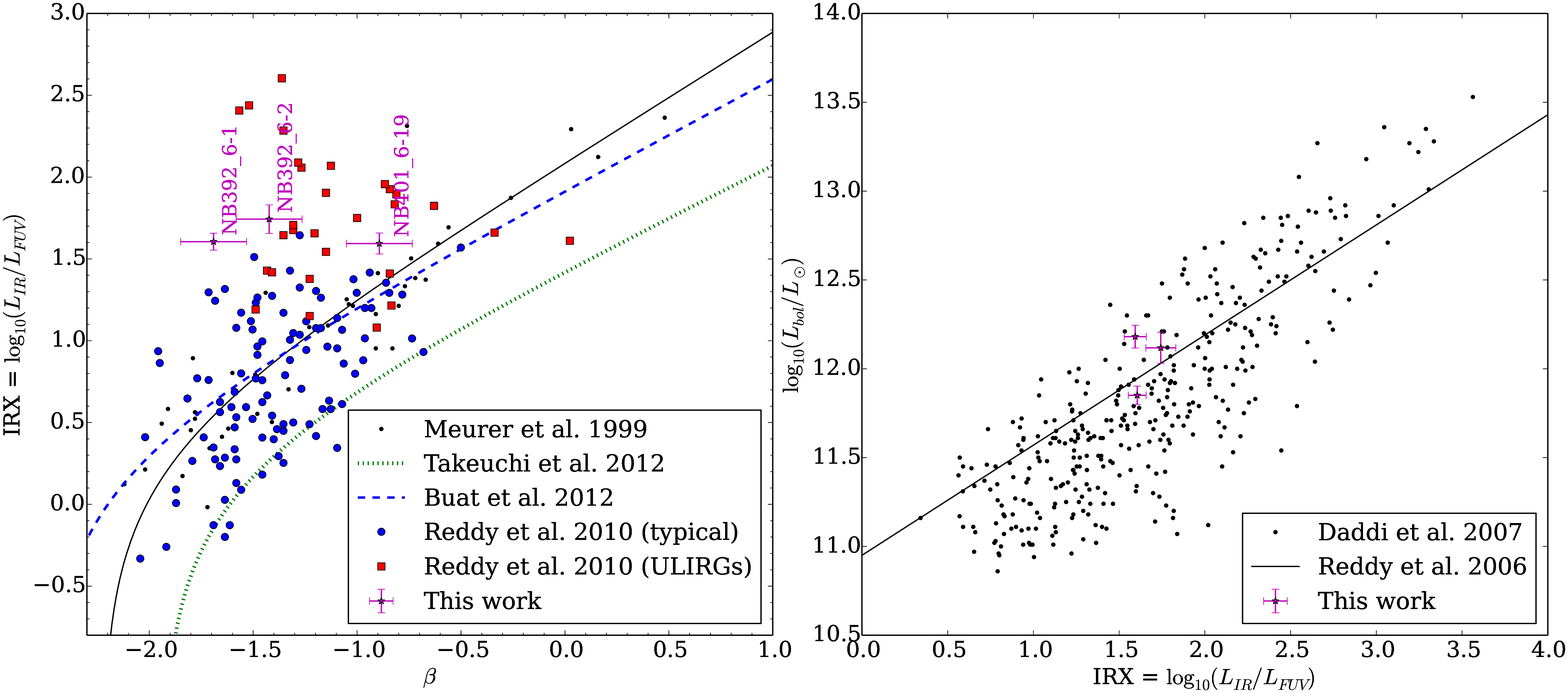}
    \caption{Left panel: $\beta$ vs. IRX plot for our three candidates detected in GOODS-Herschel. Note that the \citet{meurer-1999} data and relation have been corrected with the 1.75 factor as quoted by \citet{reddy-2010}, to convert from $L_{FIR}$ to the total $L_{IR}$. Also shown are the more recent \citet{takeuchi-2012} and \citet{buat-2012} $\beta$-IRX relations, derived for local galaxies and $z\sim1-2$ sources, respectively. Right panel: IRX vs. bolometric luminosity for the same candidates, compared to the sample of BzK selected galaxies from \citet{daddi-2007} and the empirical relation for BM/BX-selected galaxies found by \citet{reddy-2006}.}
              \label{fig:FIRIRX}%
    \end{figure*}

We have performed a match against the photometric GOODS-Herschel catalogs of \citet{elbaz-2011} and \citet{magnelli-2013}. The latter catalog combines the deeper data from \citet{elbaz-2011} with the PEP 100 and 160 $\mu$m data of \citet{lutz-2011}. 19 of our candidates are in the field of view of these surveys. 

We find five candidates with 24 $\mu$m Spitzer/MIPS detections within 1 arcsecond of the catalog coordinates. These are NB401\_6-10,-14,-19 and NB392\_6-1,-2. Three of these five candidates have detections in the \citet{magnelli-2013} PEP-Herschel catalog; NB401\_6-19, NB392\_6-1 and NB392\_6-2 (see Section~\ref{sec:SED}). 

Interestingly, NB401\_6-10 has a detection at 160 $\mu$m in the \citet{elbaz-2011} catalog, but with no detection at 100 $\mu$m and not being included in the 3$\sigma$ catalog of the combined \citet{magnelli-2013} data. It seems that the original flux may have been overestimated, and that this candidate no longer meets the significance cut in the combined catalog. At the same time, NB392\_6-1 is not listed in the \citet{elbaz-2011} catalog, but with the deeper data it is detected at 3$\sigma$.

For the three matches in \citet{magnelli-2013}, we have fitted an M82 
template to the far-IR data points, i.e. only to the datapoints from 24 $\mu$m or longer wavelengths, to derive an estimate of the total IR luminosity, $L_{IR}$, integrated from the restframe 8-1000 $\mu$m. Figure~\ref{fig:FIRfitting} shows the fit to the data, with estimates of $\log (L_{IR}/L_{\odot})$ in the bottom right corner of each graph. Also shown are two AGN templates \citep{polletta-2007} - one quasar template and a Seyfert 2 galaxy template - as well as an Arp220 template for comparison. 

Assuming the M82 template fit value as our value for $L_{IR}$, we compute the total IR to UV luminosity, IRX = $\log (L_{IR}/L_{1600})$, where $L_{1600}$ is the monochromatic luminosity at rest-frame 1600 \AA, calculated from the photometry in the V band. \citet{meurer-1999} found a tight correlation between the rest-frame UV slope $\beta$ and IRX from a local sample of starburst galaxies, indicating that the energy absorbed by dust in the UV is being proportionately re-emitted at far infrared wavelengths. \citet{reddy-2010} found that this relation breaks for ultra-luminous infrared galaxies (ULIRGs), probably due to saturation as part of the UV-bright star forming regions are being completely obscured by dust.  

We estimated the $\beta$ slopes for our three candidates by fitting a power-law through the observed V, R and I magnitudes, corresponding roughly to the 1600-2600 \AA\ range in the rest frame.
Overplotting our three candidates in the left panel of Figure~\ref{fig:FIRIRX}, they appear to fall into the same locus as the $z\sim2$ ULIRG sample from \citet{reddy-2010}. Indeed, we find $L_{IR} \gtrsim 10^{12} L_{\odot}$ for two of our three candidates, which is the very definition of a ULIRG. 
In this panel we also show the \citet{meurer-1999} relation as corrected by \citet{reddy-2010} for comparison. Note that the GALEX IRX ratios used by \citet{meurer-1999} are believed to have been overestimated \citep{overzier-2011,takeuchi-2012}. We show for comparison the new relation derived for local galaxies as presented by \citet{takeuchi-2012}, and the relation derived for $z\sim1-2$ sources from \citet{buat-2012}.
In the right panel of Figure~\ref{fig:FIRIRX} we see how our candidates compare with the IRX-luminosity ratios of $z\sim2-3$ sources from \citet{daddi-2007} and \citet{reddy-2006}. All three candidates appear to lie within the normal spread, indicating that our FIR-detected LAEs do not possess unusual UV to IR properties, compared with other samples at similar redshift. 
    
\begin{figure}
   \centering
 \includegraphics[width=\hsize]{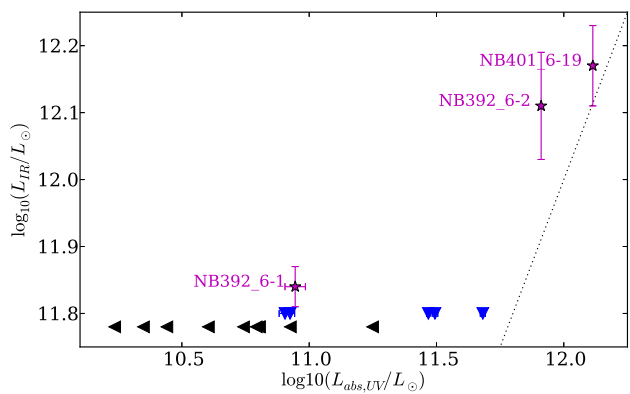}
    \caption{Far-IR luminosity versus dust-absorbed luminosity as estimated from the E(B-V) values derived from SED model A. Our three far-IR detections are marked with magenta stars. Blue triangles show estimates at our approximate detection threshold in the far-IR of $L_{IR}\sim10^{11.8} L_{\odot}$. Black triangles show upper limits to the dust absorbed luminosities, and have been shifted by -0.02 dex along the y-axis for clarity. The dotted black line shows the 1:1 relation. }
              \label{fig:LFIRvsLabs}%
    \end{figure}

We estimate the energy budget for our candidate LAEs by calculating the amount of flux absorbed by dust in the optical-UV part of the spectrum, and express it in terms of luminosity as $L_{abs,UV}$. We create model spectra based on simple SED fitting (modelling only age and E(B-V) for constant star formation rate; see model A in Section~\ref{sec:SED}), both with the best fit value of E(B-V) and with a value of 0. We then integrate both spectra for each candidate, and estimate the amount of flux absorbed by dust as the difference between the two spectra. Note that $L_{abs,UV}$ is not an observed quantity but an estimate of the amount of dust absorption, expressed in terms of the absorbed luminosity. The results are shown in Fig.~\ref{fig:LFIRvsLabs}, plotted against the estimated far-IR luminosity. Our three (U)LIRGs are found among the candidates that appear to have the highest amount of $L_{abs,UV}$, as expected. All candidates show $L_{abs,UV} < L_{IR}$, which is also expected in the scenario where a dust screen model may underestimate the total dust content. Since these estimates of $L_{abs,UV}$ should correspond to a lower limit to $L_{IR}$, we thus expect at least one of them to be bright enough to be detected as a ULIRG in the far-IR (NB401\_6-19, which is also our brightest candidate in the far-IR). A handful of the rest of the candidates would appear to be at least in the LIRG regime ($11 < \log(L_{UV,abs}) < 12$), and as discussed, two of these are also detected as LIRGs or ULIRGs in the far-IR.

\subsection{Comparison to previous work}\label{sec:previouswork}

Our findings are broadly consistent with several previous works. 
Although the LAE population is normally associated with small, dust-free systems, evidence of \lya\ emission from large, dusty galaxies does exist in the literature, but the AGN contribution to the \lya\ emissivity is not always well known. 

\citet{chapman-2005} present a spectroscopic follow-up study of 98 sub-millimeter galaxies detected at 850 $\mu$m and deep VLA 1.4 GHz data, and obtained reliable redshift estimates for 73 of these. 25 of those sources show a "strong" \lya\ line. The AGN fraction is not certain, as they cannot rule out faint AGN features, but they note that the \lya\ to [C {\sc iv}] $\lambda$ 1549 line ratios are generally consistent with what you would expect from normal starbursts. They conclude that a patchy or inhomogenous dust distribution might allow \lya\ to escape from these systems. 

\citet{casey-2012} performed a spectroscopic study of 767 galaxies individually detected in Herschel/SPIRE data. 36 of those galaxies are found to be at $z>2$, with $L_{IR} \sim 2-6 \times10^{12} L_{\odot}$, the majority of which show \lya\ in emission. 17 of the \lya\ emitting galaxies do not show any [C {\sc iv}] emission, and they estimate an AGN fraction of roughly $\sim$ 25 \%.

HST/COS spectroscopy in the ultraviolet of 11 local ULIRGs is presented in recent work by \citet{martin-2015}. 8 of the ULIRGs show \lya\ in emission, three of which appear to be AGN dominated galaxies. For the five remanining ULIRGs they estimate AGN to contribute no more than 15\% to 35\% to the bolometric luminosity. 

\citet{oteo-2012a} perform a match of 56 spectroscopically detected LAEs at $z\sim2-3.5$ in GOODS-South to the GOODS/Herschel catalogs and find 10 of them to be detected by MIPS at 24 $\mu$m, four of which have detections in the PACS-160$\mu$m band within 2 arcseconds of the optical detection. All four candidates show integrated luminosities $L_{IR} > 10^{12} L_{\sun}$, placing them in the ULIRG regime, while the other six galaxies detected in MIPS are at least in the LIRG regime ($L_{IR} > 10^{11} L_{\sun}$). Their study of 23 LAEs at $z\sim0.3$ with mid-IR/FIR detections showed only one such IR-luminous galaxy, with the rest of the sample having $L_{IR} < 10^{11} L_{\sun}$ \citep{oteo-2011,oteo-2012b}. To echo the conclusions of \citet{oteo-2012a}, this indicates that the \lya\ selection technique might trace different galaxies at different redshifts; as seen for example in GALEX LAEs which are generally not as luminous as their higher redshift counterparts \citep[see e.g.][]{cowie-2011}. 

On the other hand, our results do not appear to agree with a similar study recently performed by \citet{kusakabe-2015} on LAEs at z=2.18 in the GOODS-South field. After stacking 218 LAEs in the Herschel bands they report only an upper 3$\sigma$ limit on $L_{IR}$ of $1.1 \times 10^{10} L_{\odot}$. Our study does not include the same removal of 24 $\mu$m detections and may therefore be more vulnerable to confusion from nearby sources. However, judging from the multi-wavelength imaging we have available, our sources are well-isolated (with the possible exception of NB401\_6-19) in the Spitzer/IRAC imaging, which was used to identify detections for the Herschel sample. According to \citet{kusakabe-2015}, none of the 218 LAEs in the GOODS-South sample were detected in MIPS 24$\mu$m imaging at the 3$\sigma$ level. As previously stated, out of our 19 candidates overlapping with the MIPS footprint, five have a counterpart in the 24$\mu$m band within 1 arcsec, as reported by \citet{magnelli-2013}. 
The 218 LAEs discussed by \citet{kusakabe-2015} are however considerably fainter on average than the sample presented here (Table~\ref{tab:LandEW}); the average \lya\ luminosity is $6.6 \times 10^{41}$ erg/s, and the maximum is only $6.6 \times 10^{42}$ erg/s (Kusakabe 2014, private communication). The average luminosity for our sample is $4.2\times10^{42}$ erg/s (see Table~\ref{tab:LandEW}).

\citet{wardlow-2014} stacked three samples of 126, 280 and 92 LAEs in the Extended Chandra Deep Field-South (ECDF-S) at z$\sim$2.8, 3.1 and 4.5 respectively in Herschel/SPIRE 250, 350 and 500 $\mu$m images as well as 870 $\mu$m data from LABOCA \citep{weiss-2009}. Similarly, they put 1$\sigma$ upper limits on $L_{IR}$ of $5-8\times10^{10} L_{\odot}$, corresponding to roughly $2-3\times10^{11} L_{\odot}$ at 3$\sigma$ for all three redshift ranges. Only one of our candidates is bright enough to be detected at these longer wavelengths; NB401\_6-19 (at 250 $\mu$m), which is also the brightest in the far-IR of our candidates. 

In light of these somewhat conflicting results, it is clear that large samples of bright LAEs, in particular at the strong star-forming epoch around $z\sim2$, with matching far-IR surveys could test if this is a cosmic variance issue. 

As a final note, a preliminary analysis of our CFHT/WIRCam data of \ha\ candidates matching the redshift of our NB401\_6 \lya\ data finds both NB401\_6-19 and NB401\_6-10 also show significant \ha\ emission, which would be expected for bright infrared galaxies with high star formation rates, if the star forming regions are not too obscured by dust. The details of the full \ha\ to \lya\ match will be discussed in a forthcoming paper.

\section{SED fitting}\label{sec:SED}

We combine catalog data from \citet{capak-2004}, \citet{wang-2010} and \citet{kajisawa-2010} as described in Section \ref{sec:catalog}. In order to perform SED fitting on a relatively homogeneous and self-consistent data set, we chose to use archival Subaru+KPNO UBVRIz data rather than HST/ACS data. 
A collection of "postage stamps" from our imaging data is presented as an example in Figure~\ref{fig:cutouts}. The full set is included in the Appendix.

To perform the SED fitting, we use version 2.0 of the GalMC code \citep{acquaviva-2011}, with GALAXEV synthesis models of S. Charlot \& G. Bruzual \citep[2007, updated version of][]{bruzual-charlot2003}) and a Salpeter initial mass function (IMF). We also ran some tests, comparing the original (2003) and the more recent (2007) Charlot-Bruzual model, as well as Chabrier and Kroupa IMFs, but found no strong differences for the galaxies dominated by young stellar populations that we are mainly interested in. We use a \citet{calzetti-2000} dust absorption model. For an overview on these different models and the general effects they have on SED fitting, see the discussion in \citet{acquaviva-2011}. GalMC uses the information on the reprocessing of ionizing photons in order to add nebular emission lines, similar to the approach of \citet{schaerer-debarros2009}. The Monte Carlo Markov chains were run using 20 000 samples. The output chains were then analyzed using the April 2013 version of the GetDist program from CosmoMC \citep{lewis-bridle2002}.

For all of our SED fitting we have chosen to remove the U-band, corresponding to the spectrum blueward of \lya , from the fit. The code does include a correction for IGM absorption according to \citet{madau-1995}, but this is only an average correction. Since this parameter is not known for the individual galaxy, including the U-band could skew the results. We have also chosen to subtract the contribution of the \lya\ flux to the B-band. This is done simply by assuming the contribution from the continuum to be the same in B as in the narrowband. The uncertainty in the B-band flux was then increased by adding in quadrature the photometric uncertainty of the narrowband flux contribution. 

The metallicity is a notoriously difficult parameter to constrain in SED fitting, as it has a relatively modest effect on the overall SED shape and is also strongly degenerate with the dust reddening. We therefore chose to use a fixed metallicity of 0.2 $Z_\odot$ and only allowed the reddening to vary. We held the redshift fixed at the expected value, corresponding to \lya\ being centered in the narrowband filter. 

\subsection{Selection of SED fitting parameters}

Galaxies that we observe at $z \sim 2$ have already lived through over 2 Gyr of cosmic time. In that time, these galaxies have likely gone through several merging events with epochs of heavy star formation as well as times of more quiescent evolution. What we observe at this moment represents a snapshot in time somewhere along that evolutionary path. To accurately resolve the entiry history of such a complex system based solely on the spectral energy distribution is a challenging task, which comes down to a matter of statistics and averages. 

Based on certain assumptions, we can see which parameters would fit the observed SED the best. Rather than only assuming the standard SED fitting parameters and simply compiling our results, we wanted to see if we could say anything more definitive about the star formation histories of our LAE candidates. We therefore decided to run SED fitting using three different setups: 

\begin{itemize}

\item {\em Model A} corresponds to the common approach of fitting for the age, reddening and mass of the galaxy, using a constant star formation rate. For this model, the age values represent a form of luminosity-weighted age, based on the entire optical to near-IR SED. This age is in other words difficult to interpret in a physically meaningful way. A high (low) age is usually associated with a red (blue) spectrum, but this quantity is of course strongly degenerate with the dust reddening. In order to attempt to break some of these degeneracies, and explore the parameter space in more detail, we also attempted two more unconvential models. 

\item {\em Model B} assumes an age of the galaxy equal to 2.2 Gyr, corresponding roughly to the age of the first stars in the universe at $z \sim 2$. This is not a typical value found for the stellar model age of LAEs, but the idea of this model is to see if we can constrain a general trend in the star formation history throughout the formation of the galaxy. The star formation rate $\phi$ is then allowed to change according to a simple exponential law; $\phi = 1/|\tau| \exp(-t / \tau)$. By allowing the $\tau$ parameter to take any value $|\tau| > 1$ Myr, the star formation may either decrease or increase with time, corresponding to positive or negative $\tau$ values, respectively. The exponential models are restricted so that $|\mathrm{age}/\tau| < 13.78$.  

\item {\em Model C} assumes two separate populations, each with constant star formation histories, where one is fixed to an age of 2.2 Gyr and the other is restricted to be younger than 10 Myr. This age corresponds roughly to the maximum age for which a simple stellar population can be expected to produce ionizing continuum photons giving rise to nebular lines, including \lya . This model is thus similar to the idea in model B, where we want to see how much more recent star formation is needed in order to reproduce the observed spectrum. We can then estimate the relative contributions of the two populations to the total mass as a measurement of whether the old or young stars dominate the SED.

\end{itemize}

The idea of using these three models is not to try to find the one model that best represents the spectral shape of our data. In fact, all three models can usually find reasonable fits to all of the observed photometry. None of the models represent a truly realistic view of the star formation history and so are not meant to be compared in that sense. Instead, our idea is to try to see what SED modelling can tell us about the star formation rates in general, under certain assumptions.

\subsection{SED fitting results}

\begin{figure*}
   \centering
 \includegraphics[width=16.4cm]{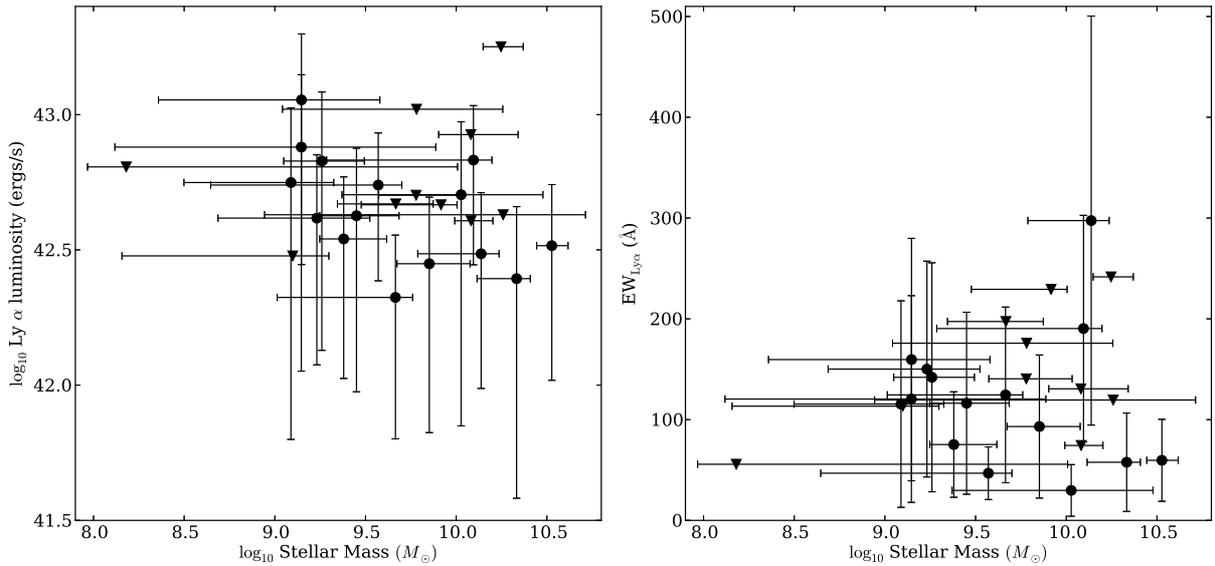}
    \caption{EW$_{\mathrm{Ly} \alpha}$ and L$_{\mathrm{Ly} \alpha}$ vs. Stellar mass, derived from SED fitting in model A. Downward pointing triangles denote upper 1$\sigma$ limits. }
              \label{fig:SmassEWlya_Llya}%
    \end{figure*}

\begin{figure*}
   \centering
 \includegraphics[width=16.4cm]{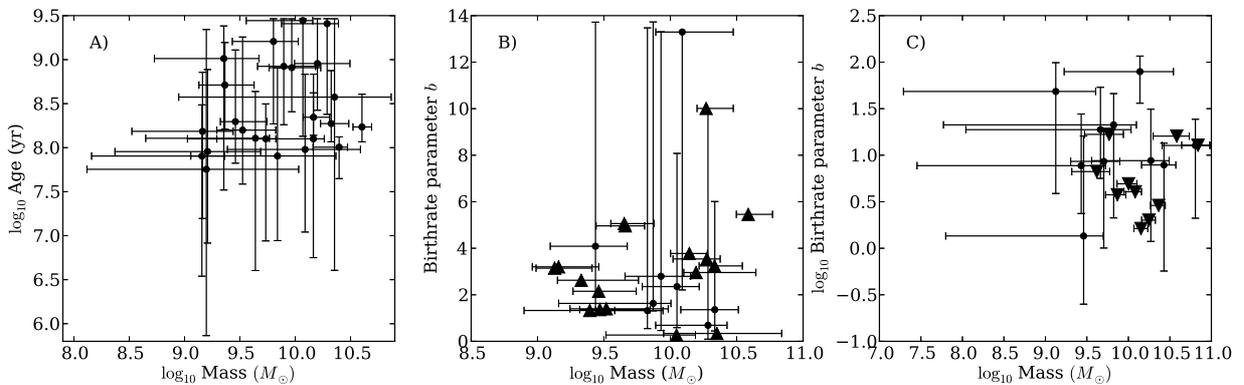}
    \caption{Left: Age vs. Mass for SED model A, using constant star formation rate. Note that NB401\_6-7 is not included in this graph as the age could not be constrained for this candidate. The mass in all three panels is the total mass that has gone into stars, i.e. the integral of the star formation rate over the age of the galaxy. Middle: Birth parameter vs. Mass for SED model B, using an exponential star formation rate (Eq. \ref{eq:bfromtau}). Lower limits in the birth parameter correspond to lower limits in 1/$\tau$ and are denoted by upward pointing triangles. Right: Birth parameter vs. Mass for SED model C. Note the logarithmic scale on the y-axis. Upper limits (1$\sigma$) are denoted with a downward pointing triangle. Four of the 25 candidates are not shown in this panel, for which the mass of the old population could not be recovered (see text). The mass is the total mass of the two populations combined.}
              \label{fig:MassvsAge}%
    \end{figure*}

The results of our fits for all our LAE candidates are presented in Tables~\ref{tab:SEDtabA}, \ref{tab:SEDtabB} and \ref{tab:SEDtabC} for models A, B and C respectively. The values given as best fit parameters are those corresponding to the lowest $\chi^2$ found. However, the probability density space may show a complicated structure, and the following "low" and "high" columns listed after each parameter represent the 68\% confidence intervals marginalized over the entire parameter space. Note that the least-squares solution may even in some cases be located outside this region. This represents a case where most reasonably good models favour certain values for that parameter (e.g., a low age), while the single model with the lowest $\chi^2$ happens to correspond to a different value for that parameter in that particular case (e.g. a high age). The marginalized limits therefore give a slightly broader picture of which parameter values can still yield acceptable fits.

The R-1 value shows the \citet{gelman-rubin1992} $R$ statistic, which is a measurement of how similar the chains are towards the final stages of the fitting. Ideally, all chains will end up in the same region of parameter space, indicating a good convergence. In cases of strong degeneracies with multiple local minima, this number may be larger, indicating that the different chains favour different solutions. R-1 $<$ 0.1 is typically considered as acceptable convergence. Any value above 0.1 is typeset in bold in Tables~\ref{tab:SEDtabA}, \ref{tab:SEDtabB} and \ref{tab:SEDtabC} for clarity. 

The general picture that emerges from the different models is not unexpected; forcing an old age generally shifts the extinction values down, making the spectrum less reddened due to dust. For the exponential star formation in model B, a negative $\tau$ value is often favored, corresponding to exponentially rising star formation, again filling in the blue spectrum by adding more newly formed stars.

In model C, we wanted to test how important recent star formation truly is for reproducing the spectrum. Our results indicate that many of these LAEs seem heavily dominated by a recent burst of star formation, while some are consistent with a slow, steady rate. Fig.~\ref{fig:MassvsAge} shows the age or the burst parameter $b$ against the mass for our three models. The $b$ parameter is defined as the current star formation rate (SFR) divided by the average past SFR. For model C, there are four candidates for which the best fit mass in the old population is negligible, corresponding to a system where a $\sim$ 10 Myr population completely dominates the spectrum. These four systems are also clearly among the lowest mass systems in model C (see Table~\ref{tab:SEDtabC}). Since we fixed the age of the old population to 2.2 Gyr, the resulting burst parameters are overly high (note the log scale in the right panel of Fig.~\ref{fig:MassvsAge}), and would for those four cases imply an essentially infinite $b$ parameter. These four candidates are therefore not shown in the figure. Note that the $b$ parameters derived for both model B and C are not to be interpreted literally, due to the heavily restricted mode of star formation in the two models. Rather, we show the $b$ parameters here as a way to understand the ratio of young to old stars in our models, and see the results only as an indication that the majority of our LAE candidates show signs of recent elevated star formation. 

For model C, the least massive systems are more inclined towards being dominated by a young population, and vice versa. When including all data points, the Pearson r coefficient in log space between the $b$ parameter and the mass is -0.659, with a Spearman rank $\rho = -0.576$. There is a tendency for this also in model A (left panel of Fig.~\ref{fig:MassvsAge}), where the ages appear slightly lower on average for the less massive systems, although the r and $\rho$ coefficients are only 0.261 and 0.303 for this case and the uncertainties are large. The trend is even less clear in model B (middle panel of Fig.~\ref{fig:MassvsAge}), where the birthrate parameter is estimated as \citep[e.g.][]{madgwick-2003}:

\begin{equation}
b = \frac{\mbox{SFR}(t)}{\left\langle \mbox{SFR}(t)\right\rangle _{\mbox{past}}}=\frac{\left(t-t_{f}\right)}{\tau}\frac{e^{-\left(t-t_{f}\right)/\tau}}{1-e^{-\left(t-t_{f}\right)/\tau}}
\label{eq:bfromtau}
\end{equation}

In eq.~\ref{eq:bfromtau}, $t-t_f$ is the difference between the current time and the time when star formation first began (i.e., the age of the galaxy), and $\tau$ is the characteristic time-scale of star formation. Note that although we can estimate the birthrate parameter using eq.~\ref{eq:bfromtau}, an exponential SFR implies a smoothly increasing or decreasing rate, rather than a sudden burst. We can thus only use model B to say whether our candidates appear to have increased or decreased their star formation rate in a more general sense.

The combined picture of these three models is clear; most, if not all, of our candidates show signatures consistent with recent, strong star formation, and the relative amount of young stars appears to be higher in the least massive systems. This is broadly consistent with a scenario where our LAEs are galaxies that have recently undergone an episode of strong star formation, across a large spread of galaxy masses. 

The \lya\ luminosities and equivalent widths are presented in Table~\ref{tab:LandEW}. These values were calculated using the continuum fit around \lya\ from model A.
Comparing our sample to similar analysis of $z\sim2$ LAEs in the literature, we can see for our model A, plotted in Fig~\ref{fig:SmassEWlya_Llya}, our sample contains relatively massive (i.e. luminous) galaxies. Compare e.g. \citet{hagen-2014}, their Fig.~7, \citet{vargas-2014}, their Fig.~5 or \citet{guaita-2011}, their Fig.~3. \citet{nilsson-2011} also found only higher stellar masses, similar to ours, for a sample of LAEs compiled in a 35 $\times$ 34 square arcminute survey using a two-meter class telescope. 

\begin{figure}
   \centering
 \includegraphics[width=\hsize]{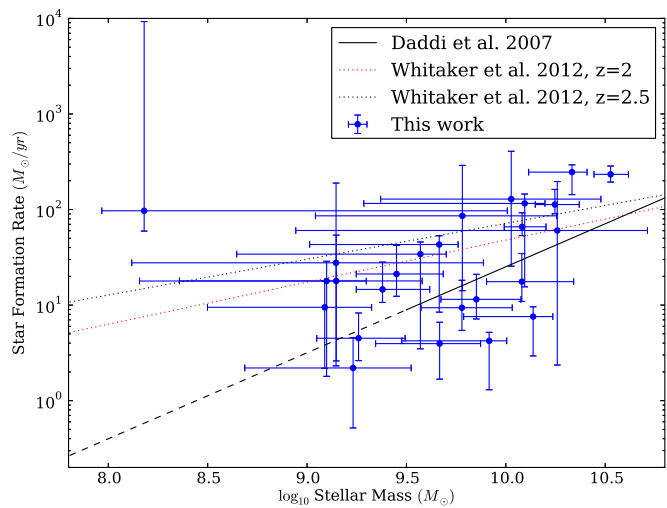}
    \caption{Stellar mass versus star formation rate for our LAEs, derived from the SED fitting in model A. The relation found for BzK selected galaxies by \citet{daddi-2007} (solid line) is also extrapolated to lower masses (dashed line). The \citet{whitaker-2012} relations for z = 2.0 and 2.5 are shown in red and black dotted lines, respectively. }
              \label{fig:SMassvsSFR}%
    \end{figure}

The average star formation rate derived from SED fitting often shows signs of being elevated in samples of LAEs, as a function of stellar mass \citep[e.g.][]{vargas-2014,hagen-2014}. This is consistent with a picture in which small, starbursting galaxies produce large amounts of ionizing photons, which are then reprocessed as \lya\, which escapes relatively easily from the lower optical depth of the smaller systems.
Looking at the star formation rate against stellar mass in Fig.~\ref{fig:SMassvsSFR}, our candidates are equally consistent with BzK-selected samples of strongly star forming galaxies \citep{daddi-2007} as with \lya -only selected galaxies at $z\sim2-2.5$ \citep{whitaker-2012}. This is not surprising given our BzK selection criterion, as well as our aim to target brighter galaxies in this survey.  

\begin{table}
\caption{\lya\ luminosity and EW}
\label{tab:LandEW}    
\centering                     
\begin{tabular}{c c c}
\hline\hline                 
Candidate & $L_{Ly~\alpha} (10^{42} \mathrm{erg s}^{-1})$ & $EW_{Ly~\alpha}$ \\
\hline
{\bf NB392\_6-1} &  $4.2 \pm 3.3$ & $116 \pm 90$ \\
{\bf NB392\_6-2} &  $6.8 \pm 4.0$ & $190 \pm 112$ \\
\hline
NB401\_6-1 &  $4.2 \pm 4.2$ & $66 \pm 65$ \\
NB401\_6-2 &  $11.3 \pm 8.5$ & $160 \pm 120$ \\
NB401\_6-3 &  $2.3 \pm 2.4$ & $112 \pm 117$ \\
NB401\_6-4 &  $5.5 \pm 3.1$ & $47 \pm 26$ \\
NB401\_6-5 &  $2.8 \pm 2.1$ & $93 \pm 71$ \\
NB401\_6-6 &  $3.1 \pm 2.1$ & $297 \pm 203$ \\
NB401\_6-7 &  $2.8 \pm 3.7$ & $24 \pm 32$ \\
NB401\_6-8 &  $4.1 \pm 3.0$ & $150 \pm 107$ \\
NB401\_6-9 &  $3.5 \pm 2.4$ & $75 \pm 52$ \\
NB401\_6-10 &  $1.9 \pm 2.1$ & $35 \pm 39$ \\
NB401\_6-11 &  $5.1 \pm 4.3$ & $30 \pm 26$ \\
NB401\_6-12 &  $5.2 \pm 5.3$ & $87 \pm 89$ \\
NB401\_6-13 &  $1.5 \pm 1.5$ & $58 \pm 55$ \\
NB401\_6-14 &  $2.5 \pm 2.1$ & $58 \pm 49$ \\
NB401\_6-15 &  $2.4 \pm 2.7$ & $66 \pm 74$ \\
NB401\_6-16 &  $5.6 \pm 5.0$ & $115 \pm 102$ \\
NB401\_6-17 &  $2.3 \pm 2.3$ & $98 \pm 99$ \\
NB401\_6-18 &  $1.9 \pm 2.3$ & $54 \pm 65$ \\
{\bf NB401\_6-19} &  $3.3 \pm 2.2$ & $60 \pm 41$ \\
NB401\_6-20 &  $2.1 \pm 1.5$ & $125 \pm 87$ \\
NB401\_6-21 &  $7.6 \pm 6.5$ & $120 \pm 103$ \\
\hline
NB406\_6-1 &  $6.7 \pm 5.4$ & $142 \pm 114$ \\
\hline
NB412\_18-1 &  $7.6 \pm 10.2$ & $103 \pm 139$ \\
\hline
\end{tabular}
\tablefoot{\lya\ luminosities and equivalent widths for our LAE candidates. These quantities were calculated using the SED fit from model A for the continuum estimation. Candidates with detections in the far-IR (Section~\ref{sec:FIR}) are marked in bold.}
\end{table}

\begin{figure}
   \centering
 \includegraphics[width=\hsize]{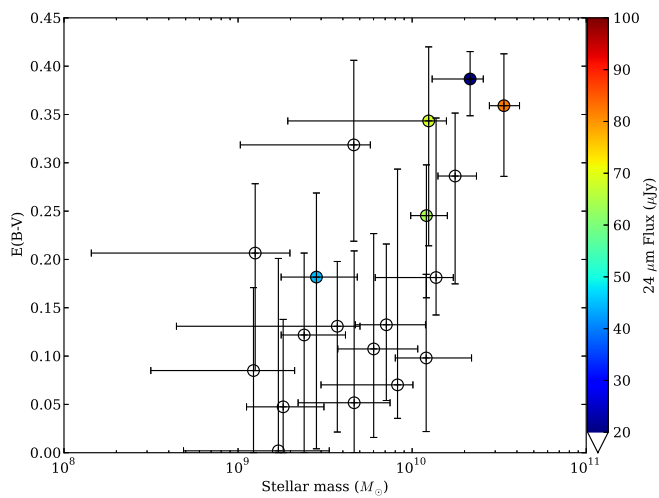}
    \caption{Stellar mass vs. E(B-V), derived from SED fit, model A. The 24 $\mu$m MIPS flux, as reported in \citet{elbaz-2011}, is indicated by color. Open circles represent no detection at 24 $\mu$m. }
              \label{fig:SmassEBV24}%
    \end{figure}

Figure~\ref{fig:SmassEBV24} shows the relationship between the mass and dust reddening as found in our SED fitting model A. The data points are marked by a colored circle, representing the amount of flux in the Spitzer/MIPS 24 $\mu$m band (see Section~\ref{sec:FIR}), as indicated by the colorbar on the right hand side. Non-detections are white in the figure. All five candidates with 24 $\mu$m detections are located in the upper half (corresponding to the highest dust obscuration) of the 19 candidates overlapping with the MIPS footprint. This indicates a strong connection between the amount of reddening of the spectrum in the optical and ultraviolet to the emission at infrared wavelengths. The 24 $\mu$m band corresponds to roughly 8 $\mu$m in the rest-frame, which is where strong PAH emission and hot dust dominates the spectrum.

\section{Conclusions}\label{sec:conclusions}

We have performed a survey of \lya\ emitting galaxies (LAEs) at $z\sim2$ in GOODS North using narrowband filters at NOT/MOSCA. Drawing upon the large quantity of ancillary data in the field, we use the entire optical to infrared catalogues to attempt to classify these LAE candidates. 

We perform a detailed SED analysis using GalMC, a Monte Carlo Markov Chain tool, using the Charlot \& Bruzual 2007 stellar population models \citep{acquaviva-2011}, testing the effects of exponential star formation histories and more than one stellar population. 

We find 25 candidates which show a wide range of ages, masses and dust reddening. Our selection is designed to catch more massive and luminous galaxies than similar recent examples in the literature \citep[e.g.][]{hagen-2014,vargas-2014,whitaker-2012,guaita-2011}, which is expected given the relatively shallow depth but large volume probed. 

We see a tendency for the least massive systems to be dominated by younger stars, and vice versa (see Fig.~\ref{fig:MassvsAge}). This is usually seen in SED modelling when assuming a constant star formation rate (as in model A), since older galaxies are forced to be more massive \citep[see e.g.][]{vargas-2014}. However, our models B and C indicate that most, if not all, of our LAE candidates show signs of recent elevated star formation, and that the light from the least massive systems is completely dominated by younger stars.

Three out of nineteen (within the footprint) LAE candidates are found in the GOODS-Herschel catalog, showing far-infrared fluxes $L_{IR} \gtrsim 10^{12} L_{\sun}$, placing them in the LIRG-ULIRG regime. An estimate of the energy absorbed in the optical-UV shows these three candidates to have among the highest $L_{UV,abs}$ of the sample (Fig.~\ref{fig:LFIRvsLabs}), as expected. We argue that these LAEs are expected to be bright far-IR emitters based on their $L_{UV,abs}$ alone, and several of our remaining LAEs may be just below the detection threshold in Herschel/PACS based on this energy budget. In conclusion, a lack of dust does not appear to be a requirement for \lya\ escape. 

\begin{acknowledgements}
The authors would like to thank Tapio Pursimo at the Nordic Optical Telescope for useful help on reducing MOSCA data and for installing and measuring our custom-built filters. We also thank Alex Hagen for fruitful discussions on SED fitting, and Viviana Acquaviva for all of her help with using and interpreting the output from GalMC. We thank Paola Santini for insightful discussions and comments on our Herschel data. 

M.H. acknowledges the support of the Swedish Research Council,
Vetenskapsr{\aa}det and the Swedish National Space Board (SNSB). 
\end{acknowledgements}

\begin{sidewaystable*}
\caption{Best fit SED parameters, model A}             
\label{tab:SEDtabA}      
\centering                          
\begin{tabular}{l l l l l l l l l l l l l l}        
\hline\hline                 
Candidate & Quality & Age (Gyr) & low & high & E(B-V) & low & high & Mass ($10^9 M_{\odot}$) & low & high & $\tilde{\chi}^2$ & d.p. & R-1 \\    

\hline                        
NB392\_6-1 & N / N / I & $0.16^{+1.65}_{-0.12}$ &  0.14 & 0.27 & $< 0.45$ & 0.00 & 0.20 & $3.3^{+3.3}_{-1.4}$ & 3.1 & 3.7 & 0.18 & 12 & 0.004 \\
NB392\_6-2 & S / I / I & $0.13^{+0.29}_{-0.12}$ &  0.05 & 0.12 & $0.34^{+0.08}_{-0.13}$ & 0.35 & 0.38 & $14.6^{+3.7}_{-12.6}$ & 9.9 & 13.6 & 0.75 & 12 & 0.008 \\
\hline
NB401\_6-1 & N / B / B & $0.90^{+2.01}_{-0.64}$ &  0.60 & 2.99 & $< 0.28$ & 0.00 & 0.12 & $15.9^{+15.4}_{-6.0}$ & 13.4 & 18.1 & 0.33 & 10 & 0.008 \\
NB401\_6-2 & N / N / I & $0.09^{+0.68}_{-0.08}$ &  0.04 & 0.12 & $< 0.28$ & 0.00 & 0.12 & $1.6^{+3.3}_{-1.4}$ & 0.8 & 1.8 & 2.15 & 6 & 0.002 \\
NB401\_6-3 & N / I / I & $2.78^{+0.14}_{-2.64}$ &  0.40 & 2.99 & $< 0.36$ & 0.00 & 0.18 & $11.7^{+2.7}_{-8.1}$ & 5.3 & 7.7 & 0.62 & 11 & $<$ 0.001 \\
NB401\_6-4 & N / I / I & $0.13^{+0.31}_{-0.12}$ &  0.05 & 0.13 & $< 0.33$ & 0.00 & 0.16 & $4.4^{+1.5}_{-3.9}$ & 2.7 & 4.2 & 2.03 & 10 & 0.048 \\
NB401\_6-5 & N / B / B & $0.81^{+2.10}_{-0.55}$ &  0.58 & 2.99 & $< 0.35$ & 0.00 & 0.15 & $9.3^{+7.7}_{-3.5}$ & 8.1 & 10.3 & 1.16 & 12 & 0.002 \\
NB401\_6-6 & S / I / I & $2.56^{+0.36}_{-2.32}$ &  0.72 & 2.99 & $0.18^{+0.17}_{-0.04}$ & 0.21 & 0.26 & $19.4^{+5.2}_{-11.9}$ & 10.7 & 15.0 & 1.39 & 12 & 0.002 \\
NB401\_6-7 & N / N / I & $0.00^{+2.79}_{-0.00}$ &  0.00 & 2.99 & $< 0.23$ & 0.00 & 0.11 & $0.15^{+14.33}_{-0.06}$ & 0.19 & 1.05 & 0.76 & 6 & 0.006 \\
NB401\_6-8 & N / B / B & $1.03^{+1.38}_{-1.00}$ &  0.17 & 0.48 & $< 0.20$ & 0.00 & 0.11 & $2.3^{+2.4}_{-1.7}$ & 1.1 & 1.8 & 0.22 & 8 & 0.005 \\
NB401\_6-9 & N / I / I & $0.20^{+1.09}_{-0.13}$ &  0.16 & 0.29 & $< 0.33$ & 0.00 & 0.14 & $2.9^{+2.7}_{-0.8}$ & 2.7 & 3.2 & 0.66 & 11 & {\bf 0.128} \\
NB401\_6-10 & N / B / I & $0.22^{+0.46}_{-0.08}$ &  0.22 & 0.33 & $0.25^{+0.05}_{-0.09}$ & 0.21 & 0.24 & $14.6^{+5.9}_{-2.8}$ & 14.3 & 16.4 & 0.37 & 12 & 0.003 \\
NB401\_6-11 & N / N / I & $0.10^{+0.59}_{-0.08}$ &  0.05 & 0.15 & $0.21^{+0.04}_{-0.08}$ & 0.19 & 0.21 & $12.3^{+26.6}_{-9.8}$ & 7.6 & 15.6 & 1.96 & 6 & 0.018 \\
NB401\_6-12 & N / N / B & $0.08^{+0.73}_{-0.07}$ &  0.04 & 0.13 & $0.27^{+0.05}_{-0.11}$ & 0.24 & 0.27 & $6.9^{+16.4}_{-5.8}$ & 4.1 & 9.1 & 1.18 & 6 & 0.009 \\
NB401\_6-13 & N / B / B & $0.08^{+0.23}_{-0.08}$ &  0.02 & 0.07 & $0.21^{+0.07}_{-0.12}$ & 0.19 & 0.22 & $1.4^{+0.9}_{-1.3}$ & 0.6 & 1.2 & 0.32 & 8 & 0.086 \\
NB401\_6-14 & N / I / I & $0.10^{+0.03}_{-0.06}$ &  0.07 & 0.10 & $0.39^{+0.03}_{-0.04}$ & 0.38 & 0.40 & $24.9^{+4.8}_{-10.4}$ & 20.1 & 23.9 & 6.21 & 12 & 0.002 \\
NB401\_6-15 & S / I / I & $0.84^{+2.07}_{-0.66}$ &  0.50 & 2.99 & $< 0.33$ & 0.00 & 0.14 & $7.9^{+7.4}_{-3.3}$ & 6.4 & 8.7 & 1.13 & 9 & 0.002 \\
NB401\_6-16 & N / N / B & $0.15^{+0.57}_{-0.14}$ &  0.07 & 0.17 & $< 0.26$ & 0.00 & 0.11 & $1.5^{+1.3}_{-1.1}$ & 0.9 & 1.5 & 1.52 & 8 & 0.004 \\
NB401\_6-17 & N / B / B & $1.61^{+1.31}_{-1.42}$ &  0.55 & 2.99 & $< 0.26$ & 0.00 & 0.12 & $6.4^{+4.3}_{-3.7}$ & 4.0 & 5.7 & 0.54 & 10 & 0.002 \\
NB401\_6-18 & N / N / I & $0.38^{+2.54}_{-0.37}$ &  0.04 & 2.99 & $0.30^{+0.11}_{-0.14}$ & 0.29 & 0.34 & $22.6^{+51.0}_{-21.8}$ & 5.7 & 20.6 & 1.46 & 6 & 0.007 \\
NB401\_6-19 & S / B / B & $0.17^{+0.23}_{-0.06}$ &  0.16 & 0.22 & $0.36^{+0.05}_{-0.07}$ & 0.33 & 0.37 & $40.1^{+8.9}_{-6.8}$ & 39.2 & 43.3 & 0.54 & 12 & 0.003 \\
NB401\_6-20 & N / B / B & $0.13^{+0.19}_{-0.12}$ &  0.04 & 0.10 & $0.32^{+0.09}_{-0.10}$ & 0.33 & 0.36 & $5.4^{+1.3}_{-4.4}$ & 3.0 & 4.7 & 0.82 & 11 & $<$ 0.001 \\
NB401\_6-21 & N / N / B & $0.06^{+2.14}_{-0.06}$ &  0.02 & 0.14 & $< 0.39$ & 0.00 & 0.16 & $1.6^{+9.2}_{-1.4}$ & 0.8 & 2.6 & 1.30 & 6 & 0.056 \\
\hline
NB406\_6-1 & N / I / I & $0.51^{+1.05}_{-0.35}$ &  0.32 & 0.58 & $< 0.19$ & 0.00 & 0.08 & $2.3^{+1.9}_{-1.0}$ & 1.9 & 2.5 & 1.29 & 11 & 0.004 \\
\hline
NB412\_18-1 & S / I / I & $0.19^{+0.56}_{-0.07}$ &  0.19 & 0.31 & $0.29^{+0.07}_{-0.11}$ & 0.24 & 0.28 & $21.1^{+9.3}_{-4.2}$ & 20.9 & 23.7 & 0.30 & 10 & 0.015 \\
\hline

\end{tabular}
\tablefoot{Derived SED properties for model A. For a detailed description on the setup of the SED fit, see the text. The Quality column lists: Spectroscopically matching LAEs at the correct redshift according to the \citet{barger-2008} or \citet{adams-2011} spectroscopic catalogs (S or N for Spectrum or No spectrum) / Possible confusion from HST imaging (B or I or N for Blended or Isolated or No data) / Possible confusion from ground-based imaging (B or I for Blended or Isolated). Mass here corresponds to the total mass that has gone into stars (i.e. the integral of the star formation rate over the age of galaxy). The R-1 value shows the \citet{gelman-rubin1992} $R$ statistic. The best fit values and their upper and lower uncertainties correspond to the least-squares values with 68\% confidence intervals. If only an upper or lower limit is listed, this corresponds to the 68\% limit. The two columns following each parameter give the 68\% lower and upper limits when marginalized over all parameters. $\tilde{\chi}^2$ represents the $\chi^2$ of the best fit model, divided by the degrees of freedom, which is the number of data points (d.p.) minus the number of parameters fitted. }
\end{sidewaystable*}

\begin{sidewaystable*}
\caption{Best fit SED parameters, model B}             
\label{tab:SEDtabB}      
\centering                          
\begin{tabular}{l l l l l l l l l l l l l l}        
\hline\hline                 

Candidate & Quality & E(B-V) & low & high & Mass ($10^9 M_{\odot}$) & low & high & 
$\tau^{-1}$ (Gyr$^{-1}$) & low & high & 
$\tilde{\chi}^2$ & d.p. & R-1 \\


\hline                        
NB392\_6-1 & N / N / I & $< 0.36$ & 0.00 & 0.15 & $2.9^{+4.1}_{-0.9}$ & 3.4 & 4.2 & $< -0.30$ & -6.34 & -1.99 & 0.30 & 12 & 0.015 \\
NB392\_6-2 & S / I / I & $0.29^{+0.02}_{-0.10}$ & 0.25 & 0.28 & $13.9^{+5.1}_{-3.9}$ & 13.4 & 15.5 & $< -1.70$ & -6.34 & -2.77 & 0.97 & 12 & $<$ 0.001 \\
\hline
NB401\_6-1 & N / B / B & $< 0.23$ & 0.00 & 0.12 & $21.6^{+10.9}_{-9.6}$ & 15.4 & 19.5 & $-0.30^{+0.97}_{-2.46}$ & -1.45 & -0.60 & 0.43 & 10 & 0.097 \\
NB401\_6-2 & N / N / I & $< 0.22$ & 0.00 & 0.08 & $2.1^{+3.6}_{-0.7}$ & 2.5 & 3.4 & $< -1.10$ & -6.34 & -2.52 & 2.45 & 6 & 0.001 \\
NB401\_6-3 & N / I / I & $< 0.28$ & 0.00 & 0.22 & $11.1^{+4.3}_{-7.9}$ & 5.6 & 7.6 & $< 1.03$ & -4.94 & -0.36 & 0.56 & 11 & $<$ 0.001 \\
NB401\_6-4 & N / I / I & $< 0.22$ & 0.00 & 0.08 & $4.6^{+1.8}_{-1.8}$ & 4.2 & 5.0 & $< -2.27$ & -6.35 & -3.25 & 2.12 & 10 & $<$ 0.001 \\
NB401\_6-5 & N / B / B & $0.13^{+0.09}_{-0.11}$ & 0.13 & 0.17 & $11.2^{+5.3}_{-5.1}$ & 8.6 & 10.8 & $-0.94^{+1.40}_{-2.78}$ & -2.87 & -0.66 & 1.26 & 12 & 0.011 \\
NB401\_6-6 & S / I / I & $< 0.51$ & 0.00 & 0.28 & $19.2^{+7.5}_{-11.5}$ & 12.0 & 18.1 & $0.33^{+1.39}_{-4.83}$ & -2.92 & 0.86 & 1.38 & 12 & 0.003 \\
NB401\_6-7 & N / N / I & $< 0.17$ & 0.00 & 0.10 & $7.4^{+2.7}_{-6.0}$ & 2.9 & 4.4 & $-0.49^{+0.23}_{-5.83}$ & -6.32 & -1.42 & 1.20 & 6 & $<$ 0.001 \\
NB401\_6-8 & N / B / B & $< 0.18$ & 0.00 & 0.10 & $2.5^{+1.3}_{-1.7}$ & 1.4 & 1.9 & $< -0.27$ & -5.45 & -1.67 & 0.23 & 8 & 0.008 \\
NB401\_6-9 & N / I / I & $< 0.29$ & 0.00 & 0.12 & $2.9^{+2.6}_{-1.0}$ & 3.1 & 3.7 & $< -0.83$ & -6.34 & -2.48 & 0.65 & 11 & 0.002 \\
NB401\_6-10 & N / B / I & $0.24^{+0.04}_{-0.06}$ & 0.23 & 0.25 & $18.8^{+4.9}_{-8.4}$ & 15.4 & 18.3 & $-4.10^{+2.52}_{-2.24}$ & -6.34 & -2.59 & 0.37 & 12 & $<$ 0.001 \\
NB401\_6-11 & N / N / I & $0.19^{+0.03}_{-0.06}$ & 0.17 & 0.19 & $15.6^{+28.5}_{-3.0}$ & 20.1 & 26.6 & $< -1.28$ & -6.33 & -2.46 & 2.17 & 6 & $<$ 0.001 \\
NB401\_6-12 & N / N / B & $0.23^{+0.04}_{-0.08}$ & 0.20 & 0.23 & $12.3^{+17.5}_{-4.5}$ & 13.9 & 18.6 & $-6.13^{+5.27}_{-0.00}$ & -4.50 & -1.32 & 1.55 & 6 & $<$ 0.001 \\
NB401\_6-13 & N / B / B & $< 0.32$ & 0.00 & 0.13 & $1.3^{+1.2}_{-0.4}$ & 1.5 & 1.9 & $< -1.38$ & -5.83 & -2.84 & 0.65 & 8 & $<$ 0.001 \\
NB401\_6-14 & N / I / I & $0.32^{+0.02}_{-0.03}$ & 0.30 & 0.32 & $18.6^{+11.2}_{-2.7}$ & 21.1 & 24.0 & $< -4.61$ & -6.35 & -4.94 & 7.65 & 12 & 0.023 \\
NB401\_6-15 & S / I / I & $0.12^{+0.11}_{-0.12}$ & 0.13 & 0.18 & $8.5^{+5.7}_{-4.0}$ & 6.5 & 8.4 & $-1.19^{+1.83}_{-4.94}$ & -4.36 & -0.97 & 1.16 & 9 & 0.009 \\
NB401\_6-16 & N / N / B & $< 0.19$ & 0.00 & 0.07 & $1.4^{+1.5}_{-0.5}$ & 1.5 & 2.0 & $< -1.40$ & -5.82 & -3.06 & 1.56 & 8 & 0.002 \\
NB401\_6-17 & N / B / B & $< 0.27$ & 0.00 & 0.16 & $6.7^{+2.1}_{-4.0}$ & 4.0 & 5.3 & $-0.27^{+0.78}_{-5.94}$ & -4.26 & -0.71 & 0.54 & 10 & $<$ 0.001 \\
NB401\_6-18 & N / N / I & $0.31^{+0.07}_{-0.20}$ & 0.27 & 0.32 & $22.4^{+46.6}_{-13.5}$ & 18.7 & 27.7 & $< 0.87$ & -5.29 & -1.02 & 1.46 & 6 & 0.005 \\
NB401\_6-19 & S / B / B & $0.36^{+0.02}_{-0.06}$ & 0.33 & 0.35 & $38.6^{+20.2}_{-7.2}$ & 41.7 & 48.3 & $< -2.50$ & -6.34 & -3.48 & 0.56 & 12 & $<$ 0.001 \\
NB401\_6-20 & N / B / B & $0.27^{+0.03}_{-0.07}$ & 0.24 & 0.27 & $4.5^{+3.0}_{-0.9}$ & 5.0 & 5.9 & $< -2.32$ & -6.34 & -3.21 & 0.94 & 11 & $<$ 0.001 \\
NB401\_6-21 & N / N / B & $< 0.31$ & 0.00 & 0.13 & $3.3^{+6.3}_{-1.5}$ & 3.5 & 5.0 & $< -0.33$ & -5.57 & -1.86 & 1.40 & 6 & $<$ 0.001 \\
\hline
NB406\_6-1 & N / I / I & $< 0.19$ & 0.00 & 0.09 & $2.7^{+2.0}_{-1.5}$ & 2.0 & 2.7 & $-1.85^{+1.57}_{-4.47}$ & -4.97 & -1.41 & 1.34 & 11 & 0.013 \\
\hline
NB412\_18-1 & S / I / I & $0.29^{+0.03}_{-0.09}$ & 0.25 & 0.28 & $21.5^{+13.4}_{-5.0}$ & 22.4 & 26.4 & $< -1.43$ & -5.72 & -2.61 & 0.32 & 10 & 0.003 \\
\hline

\end{tabular}
\tablefoot{Same as Table~\ref{tab:SEDtabA}, but for model B. The Age of each candidate is fixed at 2.2 Gyr.}
\end{sidewaystable*}

\begin{sidewaystable*}
\caption{Best fit SED parameters, model C}             
\label{tab:SEDtabC}      
\centering                          
\begin{tabular}{l l l l l l l l l l l l l l l l l l}
\hline\hline                 

Candidate & Quality & Age (Myr) & low & high & E(B-V) & low & high & M$_{tot} / 10^9 M_{\odot}$ & low & high & f$_{young}$ (\%) & low & high & $\tilde{\chi}^2$ & d.p. & R-1 \\


\hline                        
NB392\_6-1 & N / N / I & $6.94^{+2.86}_{-6.21}$ & 0.73 & 10.02 & $< 0.33$ & 0.00 & 0.09 & $5.9^{+2.8}_{-2.9}$ & 5.4 & 6.9 & $< 7.5$ & 0.8 & 2.1 & 0.36 & 12 & 0.020 \\
NB392\_6-2 & S / I / I & $4.31^{+5.50}_{-3.58}$ & 0.73 & 10.02 & $0.26^{+0.11}_{-0.13}$ & 0.21 & 0.26 & $18.6^{+12.4}_{-15.0}$ & 15.9 & 21.8 & $3.9^{+10.0}_{-3.3}$ & 2.6 & 4.6 & 0.83 & 12 & 0.001 \\
\hline
NB401\_6-1 & N / B / B & $4.25^{+5.57}_{-3.52}$ & 0.73 & 10.02 & $< 0.18$ & 0.00 & 0.06 & $23.1^{+4.7}_{-4.7}$ & 21.3 & 23.8 & $< 1.3$ & 0.0 & 0.1 & 0.49 & 10 & 0.007 \\
NB401\_6-2 & N / N / I & $9.61^{+0.21}_{-6.51}$ & 5.54 & 10.02 & $< 0.23$ & 0.00 & 0.08 & $0.27^{+6.44}_{-0.20}$ & 0.16 & 0.28 & $> 24.3$ & 58.4 & 81.6 & 4.66 & 6 & 0.001 \\
NB401\_6-3 & N / I / I & $0.82^{+9.00}_{-0.09}$ & 0.73 & 10.02 & $< 0.28$ & 0.00 & 0.09 & $10.0^{+2.6}_{-2.8}$ & 9.1 & 10.5 & $< 2.3$ & 0.0 & 0.1 & 0.68 & 11 & $<$ 0.001 \\
NB401\_6-4 & N / I / I & $4.63^{+5.16}_{-3.90}$ & 0.73 & 10.02 & $< 0.29$ & 0.00 & 0.14 & $4.6^{+6.2}_{-4.5}$ & 0.4 & 4.5 & $8.0^{+14.8}_{-5.6}$ & 8.1 & 12.3 & 2.24 & 10 & 0.003 \\
NB401\_6-5 & N / B / B & $7.46^{+2.35}_{-6.73}$ & 0.73 & 10.02 & $< 0.21$ & 0.00 & 0.08 & $14.2^{+2.9}_{-2.5}$ & 13.4 & 14.9 & $< 0.8$ & 0.0 & 0.0 & 1.59 & 12 & 0.002 \\
NB401\_6-6 & S / I / I & $7.40^{+2.42}_{-6.67}$ & 0.73 & 10.02 & $0.19^{+0.08}_{-0.03}$ & 0.18 & 0.20 & $17.7^{+3.4}_{-3.2}$ & 16.4 & 18.4 & $< 0.9$ & 0.0 & 0.0 & 1.57 & 12 & $<$ 0.001 \\
NB401\_6-7 & N / N / I & $1.23^{+8.59}_{-0.50}$ & 0.73 & 10.02 & $< 0.20$ & 0.00 & 0.08 & $0.15^{+11.09}_{-0.13}$ & 0.13 & 0.34 & $> 17.3$ & 88.1 & 100.0 & 1.13 & 6 & 0.003 \\
NB401\_6-8 & N / B / B & $5.38^{+4.44}_{-4.65}$ & 0.73 & 10.02 & $< 0.18$ & 0.00 & 0.07 & $2.9^{+2.1}_{-2.8}$ & 1.3 & 2.7 & $0.6^{+6.7}_{-0.5}$ & 1.2 & 2.3 & 0.29 & 8 & 0.002 \\
NB401\_6-9 & N / I / I & $8.77^{+1.04}_{-8.05}$ & 0.73 & 10.02 & $< 0.28$ & 0.00 & 0.09 & $5.1^{+2.8}_{-3.0}$ & 4.3 & 5.7 & $3.8^{+4.4}_{-3.4}$ & 1.6 & 2.9 & 0.74 & 11 & $<$ 0.001 \\
NB401\_6-10 & N / B / I & $9.74^{+0.08}_{-9.01}$ & 0.73 & 10.02 & $0.23^{+0.05}_{-0.12}$ & 0.16 & 0.20 & $26.8^{+10.5}_{-12.4}$ & 25.6 & 29.9 & $3.5^{+2.5}_{-3.2}$ & 1.1 & 2.0 & 0.69 & 12 & 0.014 \\
NB401\_6-11 & N / N / I & $9.81^{+0.01}_{-9.07}$ & 5.99 & 10.02 & $< 0.45$ & 0.00 & 0.20 & $2.4^{+76.0}_{-2.4}$ & 1.4 & 8.8 & ... & 57.2 & 83.6 & 4.67 & 6 & 0.001 \\
NB401\_6-12 & N / N / B & $9.80^{+0.02}_{-6.60}$ & 5.71 & 10.02 & $0.25^{+0.04}_{-0.08}$ & 0.21 & 0.24 & $1.4^{+25.4}_{-0.9}$ & 0.8 & 1.2 & $> 35.5$ & 56.5 & 79.2 & 3.24 & 6 & $<$ 0.001 \\
NB401\_6-13 & N / B / B & $9.03^{+0.79}_{-8.30}$ & 0.73 & 10.02 & $< 0.45$ & 0.00 & 0.18 & $1.3^{+2.7}_{-1.3}$ & 0.1 & 0.9 & $18.3^{+18.9}_{-16.8}$ & 10.4 & 15.7 & 0.54 & 8 & 0.002 \\
NB401\_6-14 & N / I / I & $9.81^{+0.01}_{-4.59}$ & 7.61 & 10.02 & $0.37^{+0.05}_{-0.06}$ & 0.37 & 0.38 & $13.8^{+21.1}_{-12.1}$ & 3.3 & 4.9 & $26.7^{+12.6}_{-14.5}$ & 24.3 & 29.8 & 8.88 & 12 & $<$ 0.001 \\
NB401\_6-15 & S / I / I & $9.74^{+0.08}_{-9.01}$ & 0.73 & 10.02 & $< 0.25$ & 0.00 & 0.07 & $12.1^{+2.3}_{-3.2}$ & 10.9 & 12.3 & $< 1.9$ & 0.0 & 0.0 & 1.52 & 9 & $<$ 0.001 \\
NB401\_6-16 & N / N / B & $9.23^{+0.59}_{-8.50}$ & 4.42 & 10.02 & $< 0.20$ & 0.00 & 0.07 & $2.7^{+2.6}_{-2.7}$ & 0.1 & 1.6 & $3.4^{+8.9}_{-2.4}$ & 3.4 & 5.2 & 2.31 & 8 & $<$ 0.001 \\
NB401\_6-17 & N / B / B & $0.96^{+8.86}_{-0.23}$ & 0.73 & 10.02 & $< 0.19$ & 0.00 & 0.06 & $7.4^{+1.9}_{-2.1}$ & 6.6 & 7.7 & $< 1.7$ & 0.0 & 0.1 & 0.64 & 10 & 0.004 \\
NB401\_6-18 & N / N / I & $9.25^{+0.57}_{-8.52}$ & 0.73 & 10.02 & $0.34^{+0.04}_{-0.19}$ & 0.23 & 0.29 & $1.9^{+74.5}_{-1.9}$ & 0.3 & 34.5 & ... & 17.9 & 56.3 & 2.52 & 6 & 0.002 \\
NB401\_6-19 & S / B / B & $9.44^{+0.38}_{-8.71}$ & 0.73 & 10.02 & $0.35^{+0.06}_{-0.12}$ & 0.29 & 0.33 & $64.1^{+34.6}_{-38.7}$ & 58.7 & 74.8 & $5.5^{+5.1}_{-4.6}$ & 2.4 & 4.2 & 0.82 & 12 & 0.004 \\
NB401\_6-20 & N / B / B & $9.28^{+0.54}_{-8.55}$ & 0.73 & 10.02 & $0.30^{+0.08}_{-0.16}$ & 0.23 & 0.30 & $6.6^{+5.8}_{-6.6}$ & 3.7 & 7.5 & $8.9^{+10.4}_{-8.0}$ & 4.1 & 8.4 & 0.86 & 11 & 0.003 \\
NB401\_6-21 & N / N / B & $9.45^{+0.37}_{-8.71}$ & 0.73 & 10.02 & $< 0.34$ & 0.00 & 0.10 & $0.37^{+10.64}_{-0.33}$ & 0.15 & 0.39 & $> 10.3$ & 41.6 & 60.1 & 2.25 & 6 & 0.004 \\
\hline
NB406\_6-1 & N / I / I & $9.18^{+0.64}_{-8.45}$ & 0.73 & 10.02 & $< 0.11$ & 0.00 & 0.02 & $4.2^{+1.8}_{-2.1}$ & 3.8 & 4.9 & $< 3.1$ & 0.3 & 0.8 & 1.82 & 11 & 0.001 \\
\hline
NB412\_18-1 & S / I / I & $9.52^{+0.30}_{-8.79}$ & 0.73 & 10.02 & $0.27^{+0.06}_{-0.15}$ & 0.18 & 0.23 & $38.2^{+15.9}_{-18.4}$ & 35.4 & 43.1 & $3.9^{+3.2}_{-3.8}$ & 1.0 & 2.4 & 0.62 & 10 & $<$ 0.001 \\
\hline

\end{tabular}
\tablefoot{Same as Table~\ref{tab:SEDtabA}, but for model C. f$_{young}$ represents the fraction of the mass found in the younger population. A missing value means the fraction could not be constrained.}
\end{sidewaystable*}


\bibliographystyle{aa} 
\bibliography{sandbergetal2013.bib} 

\Online

\begin{appendix} 
\section{Postage stamps of all candidate LAEs}

\begin{figure*}
\centering
\includegraphics{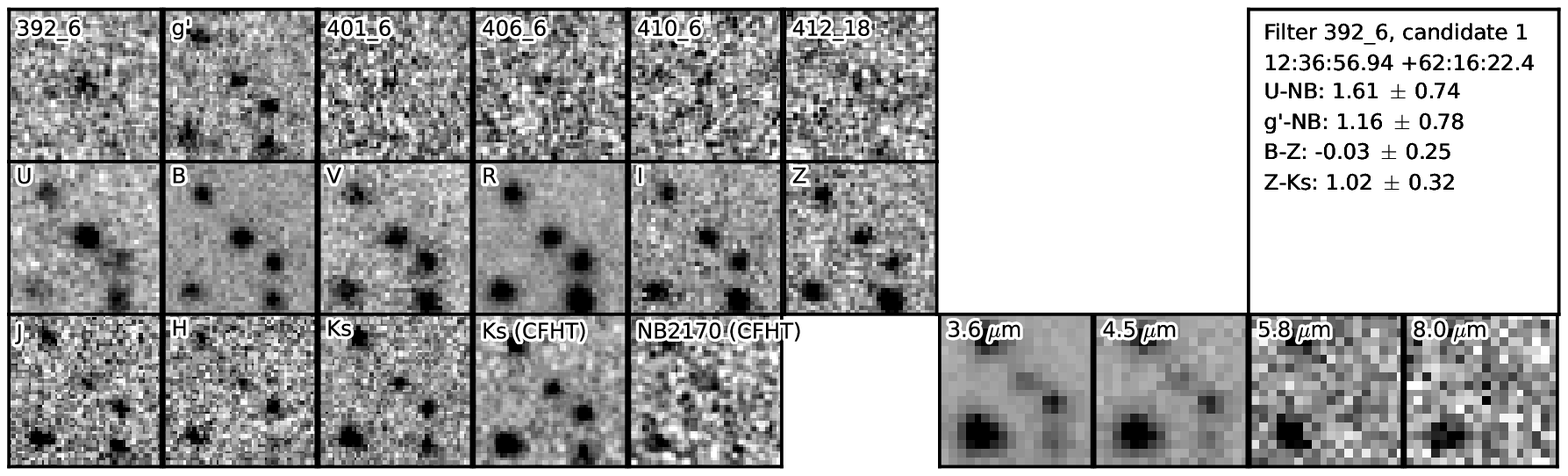}
\includegraphics{ALLstamps/stamp392_6-02.eps}
\label{fig:allstamps}
\caption{Postage stamps for all of our candidate LAEs. The field of view of all stamps is 10 $\times$ 10 arcsec. The large RGB picture shows an HST $bvi$ color composite made with the scale in each color channel related to the true total intensity, with an arcsinh intensity scaling relation. This coloring scheme is similar to that described in \citet{lupton-2004}. The white circle has a one arcsec radius - typically twice the size of the ground-based seeing disk. }
\end{figure*}

We here show the postage stamps for all of our candidates in Figure~\ref{fig:allstamps}. 

\begin{figure*}
\centering
\includegraphics{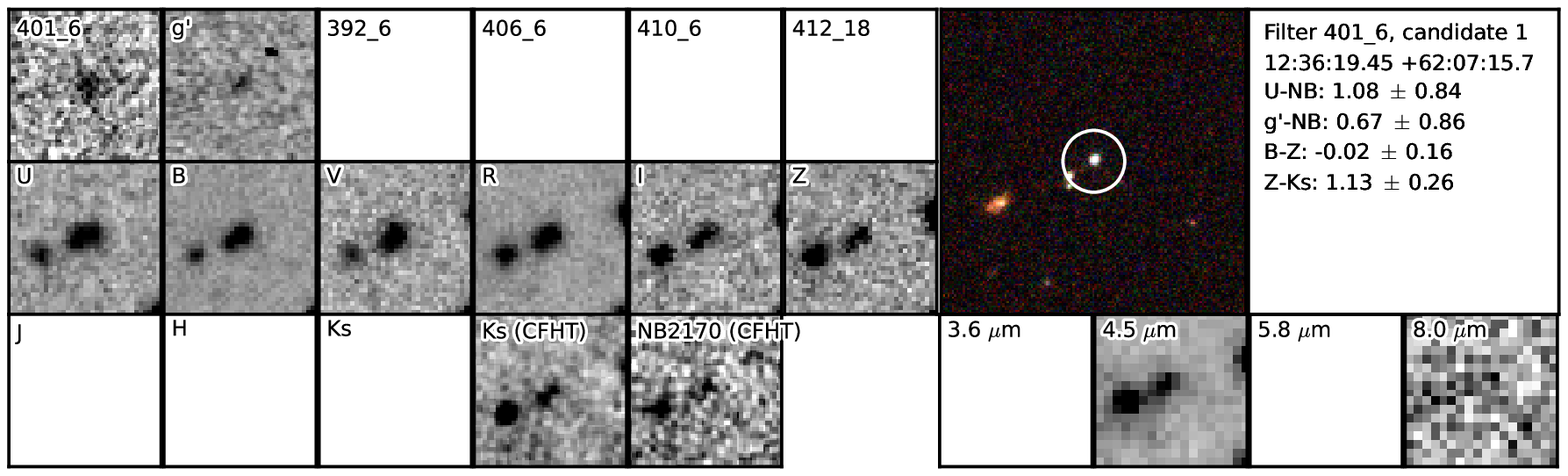}
\includegraphics{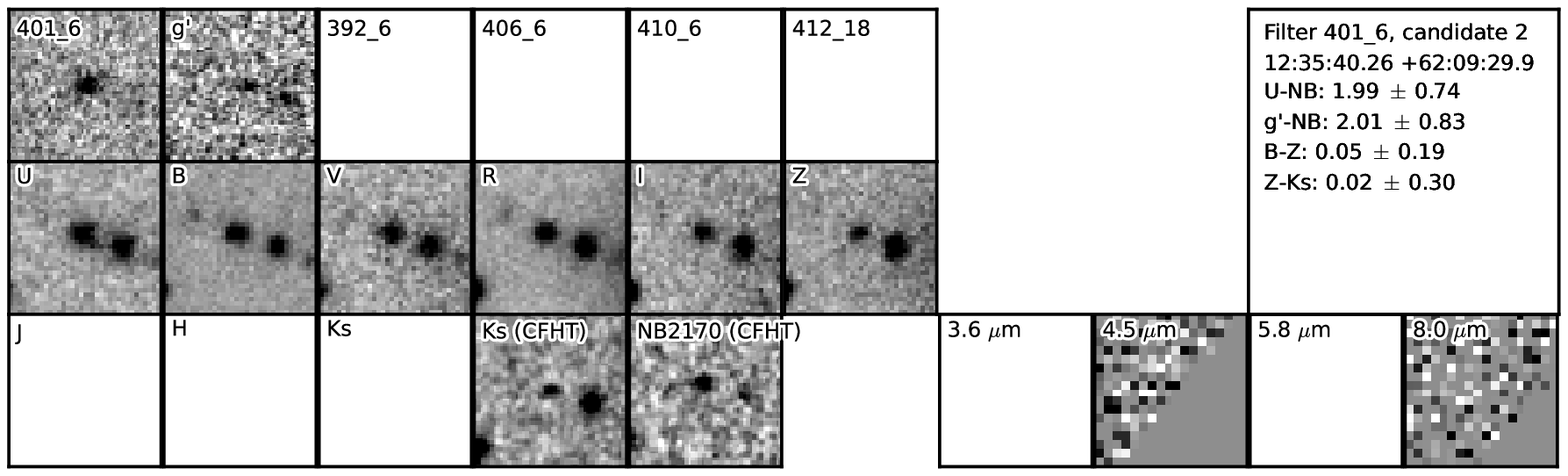}
\includegraphics{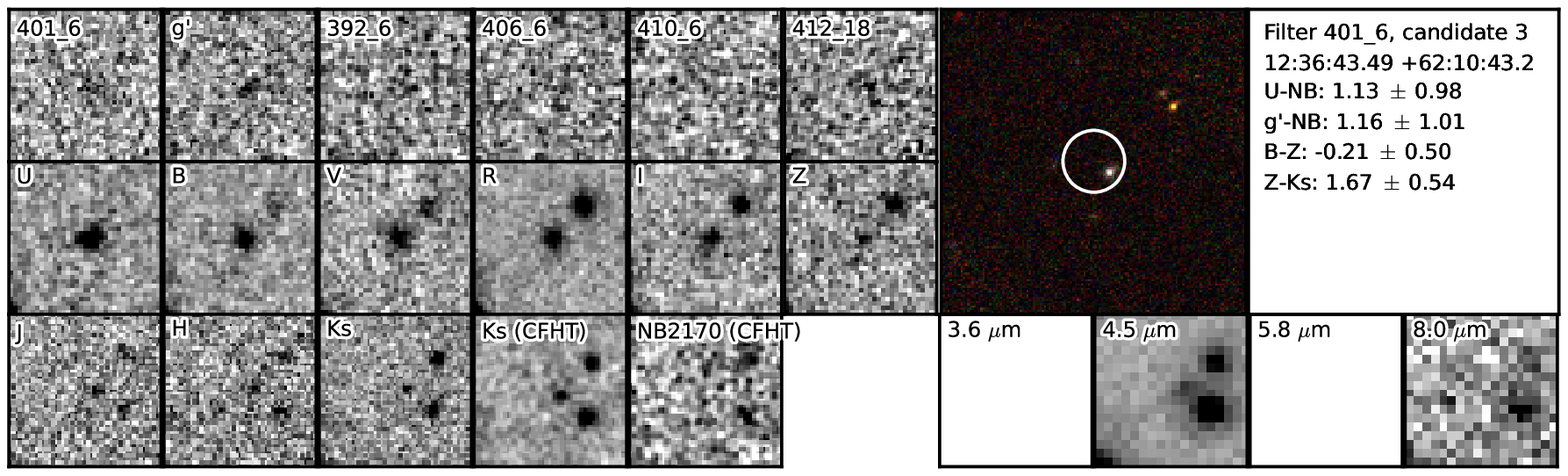}
\includegraphics{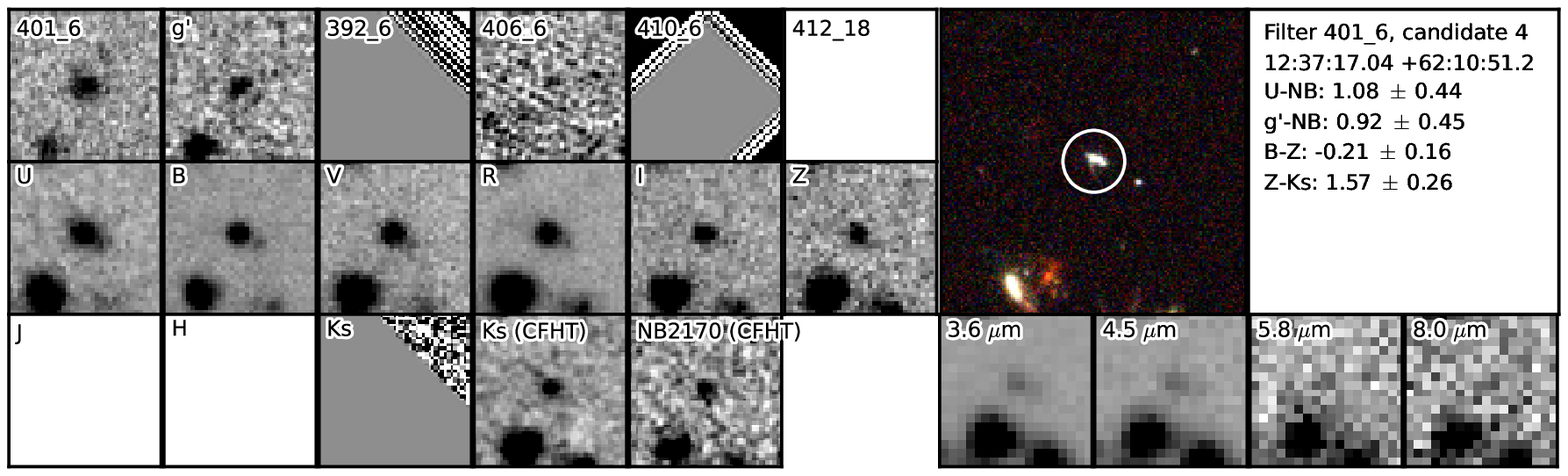}
\caption{Stamps (cont.)}
\end{figure*}

\begin{figure*}
\centering
\includegraphics{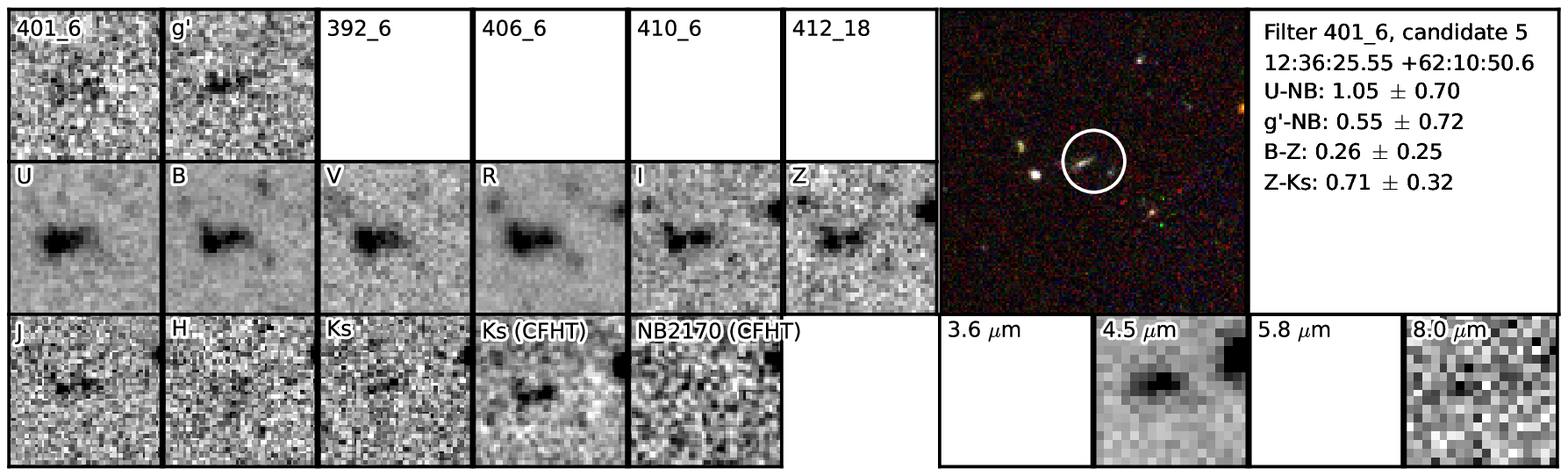}
\includegraphics{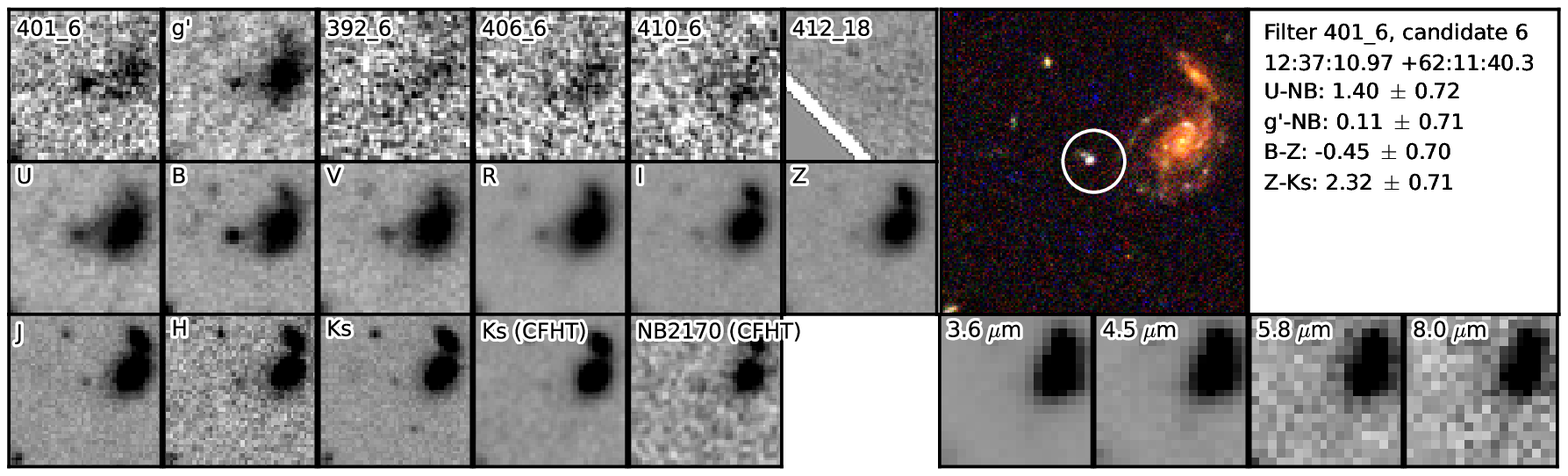}
\includegraphics{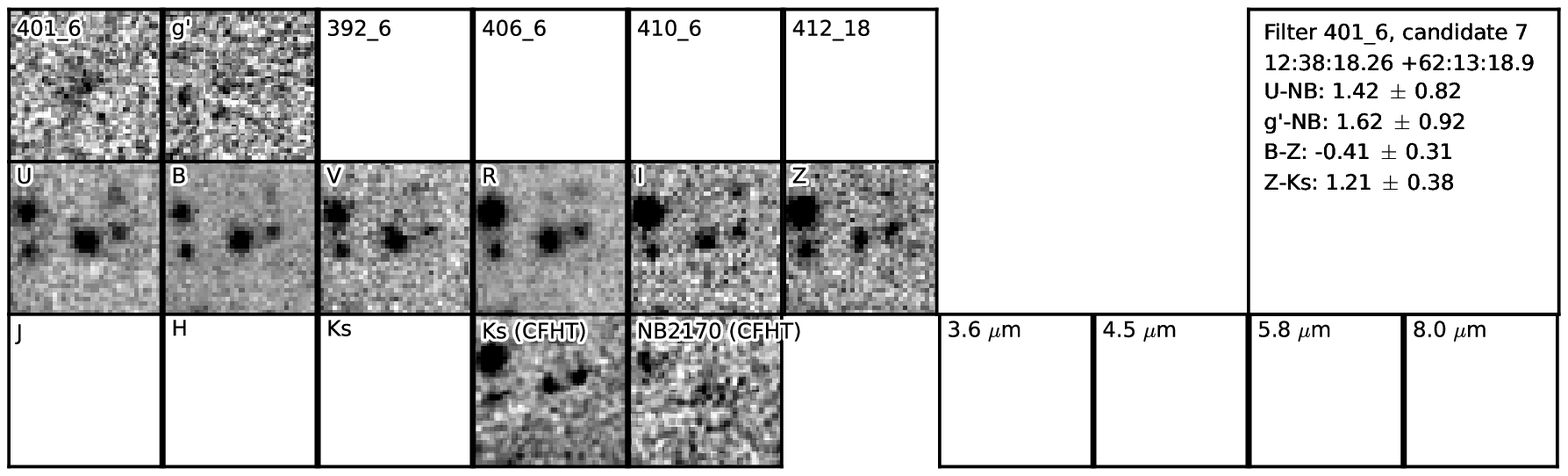}
\includegraphics{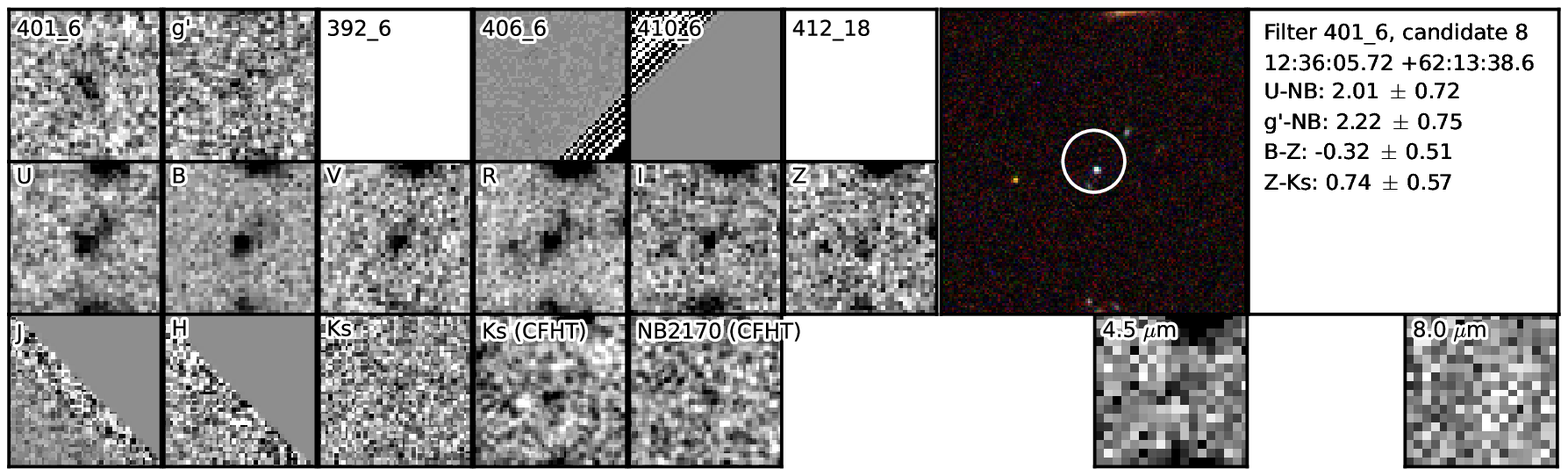}
\caption{Stamps (cont.)}
\end{figure*}

\begin{figure*}
\centering
\includegraphics{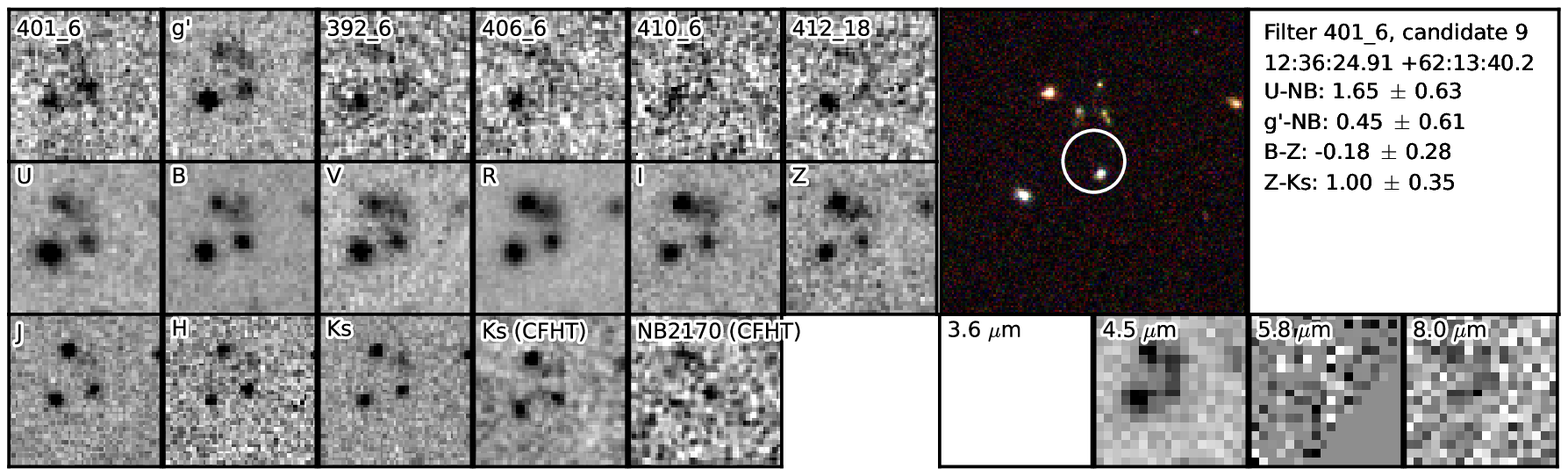}
\includegraphics{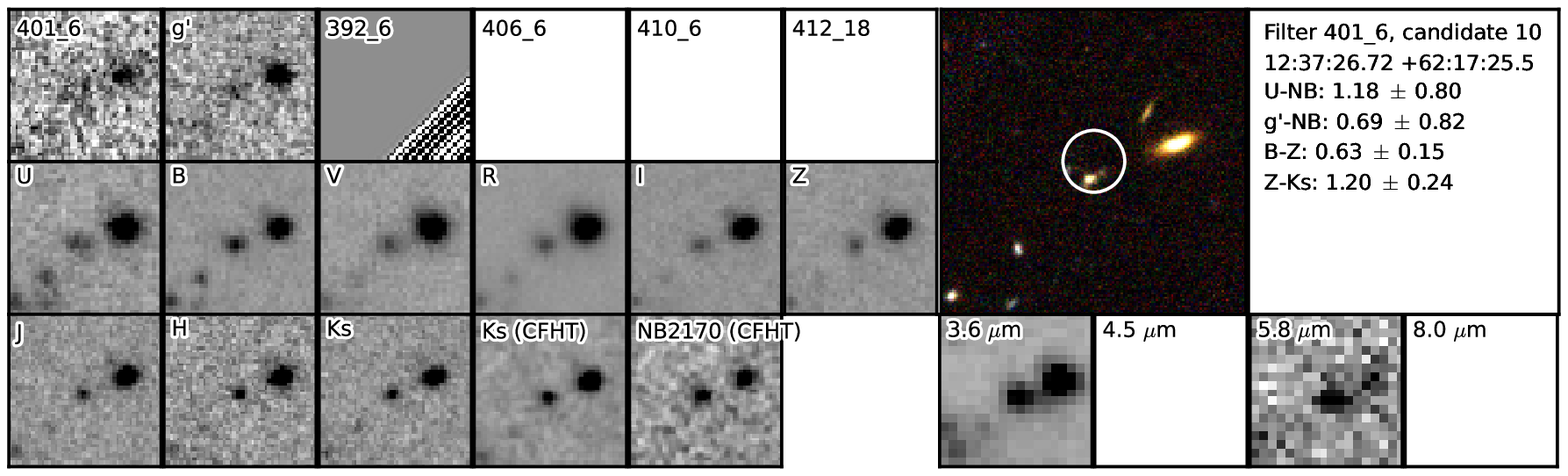}
\includegraphics{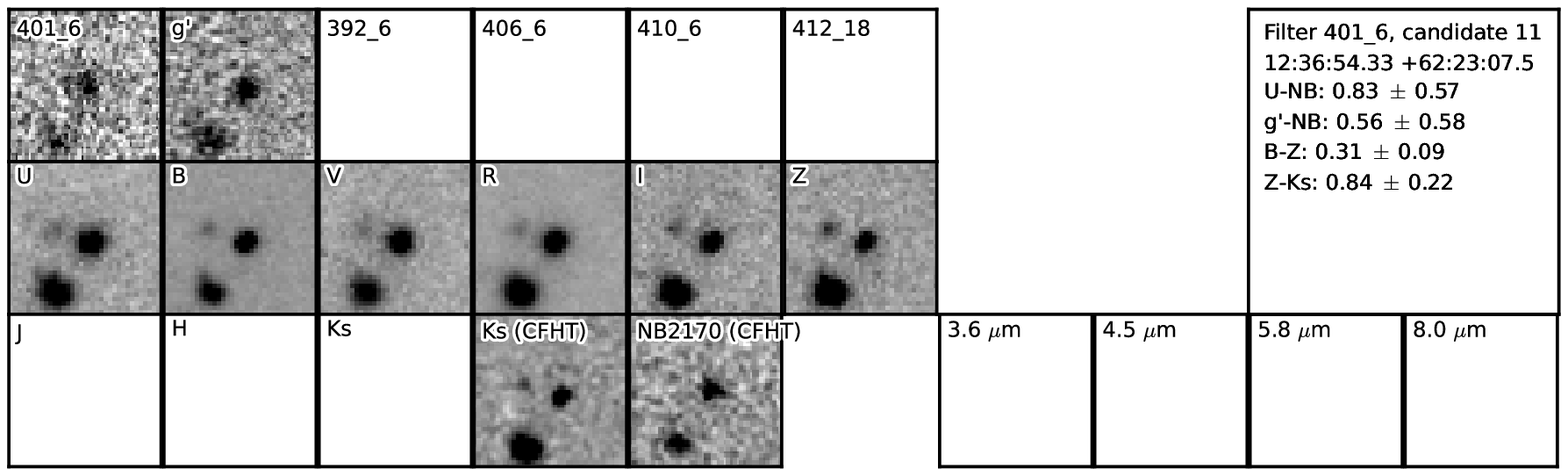}
\includegraphics{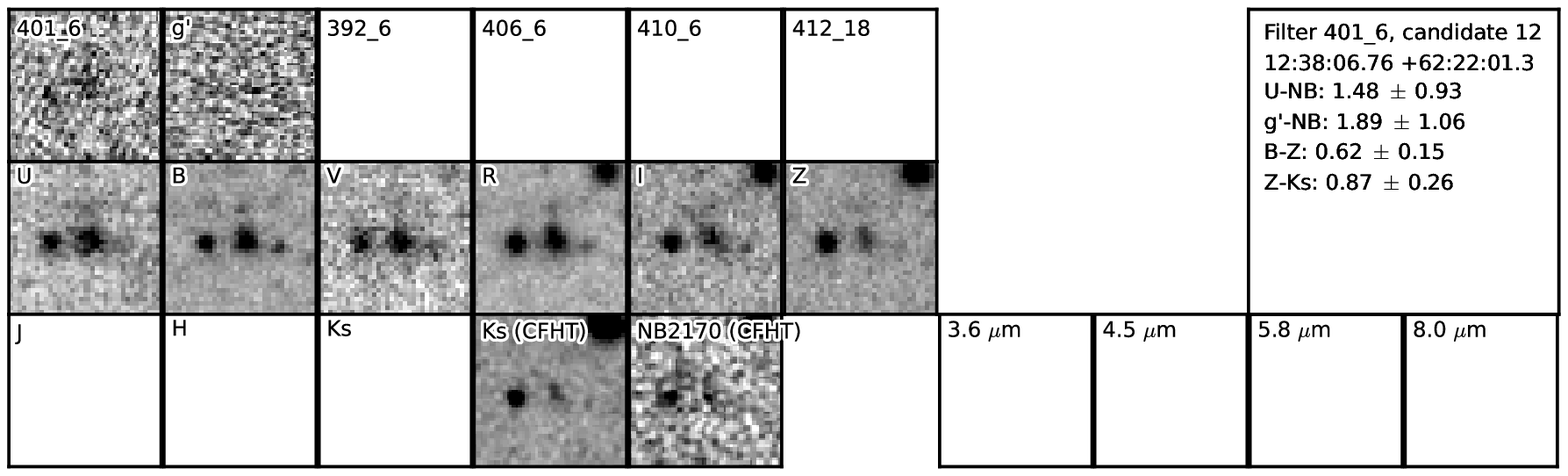}
\caption{Stamps (cont.)}
\end{figure*}

\begin{figure*}
\centering
\includegraphics{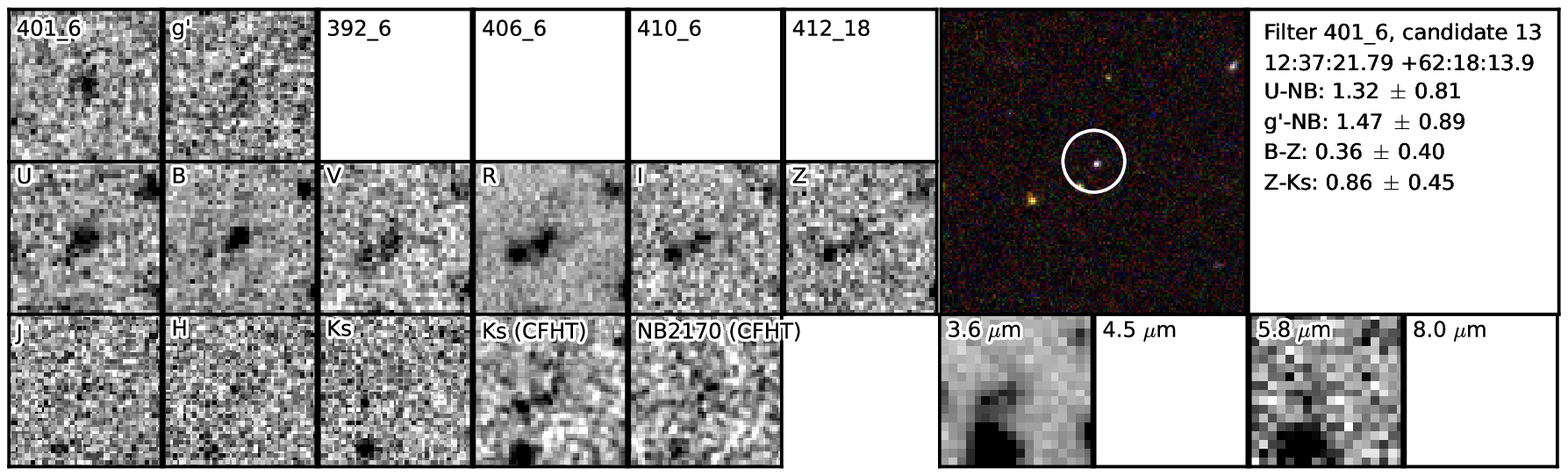}
\includegraphics{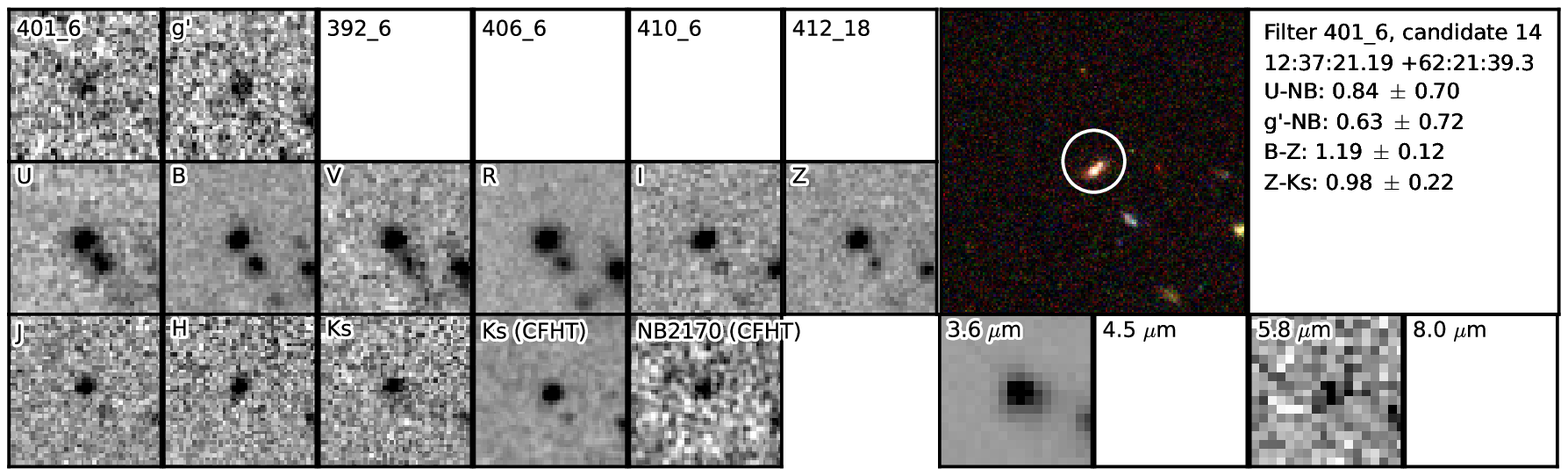}
\includegraphics{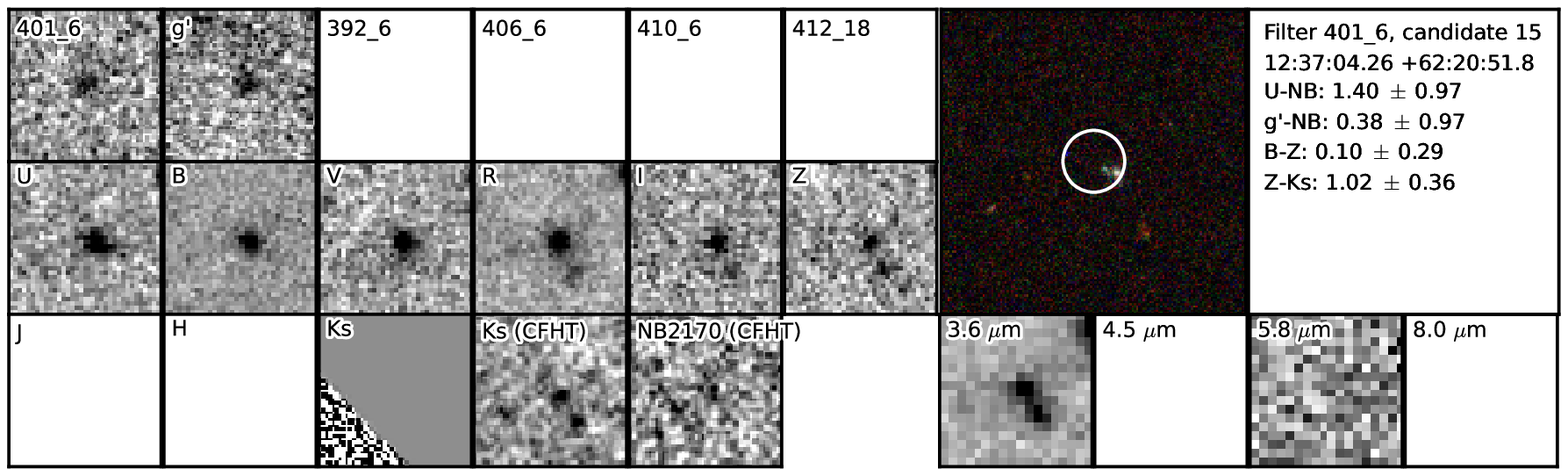}
\includegraphics{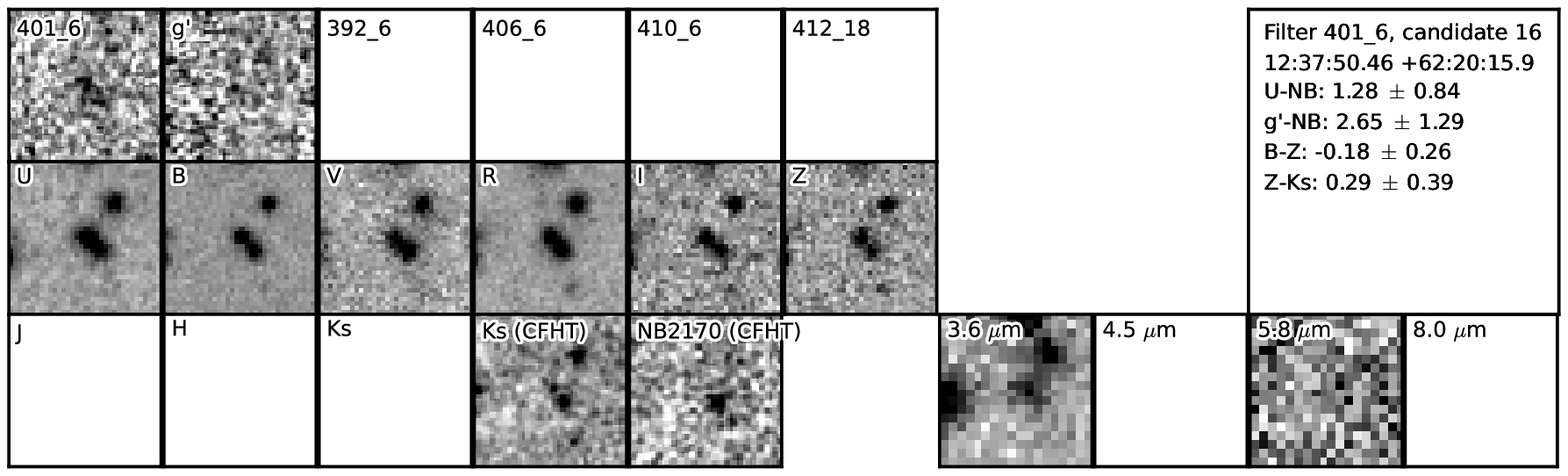}
\caption{Stamps (cont.)}
\end{figure*}

\begin{figure*}
\centering
\includegraphics{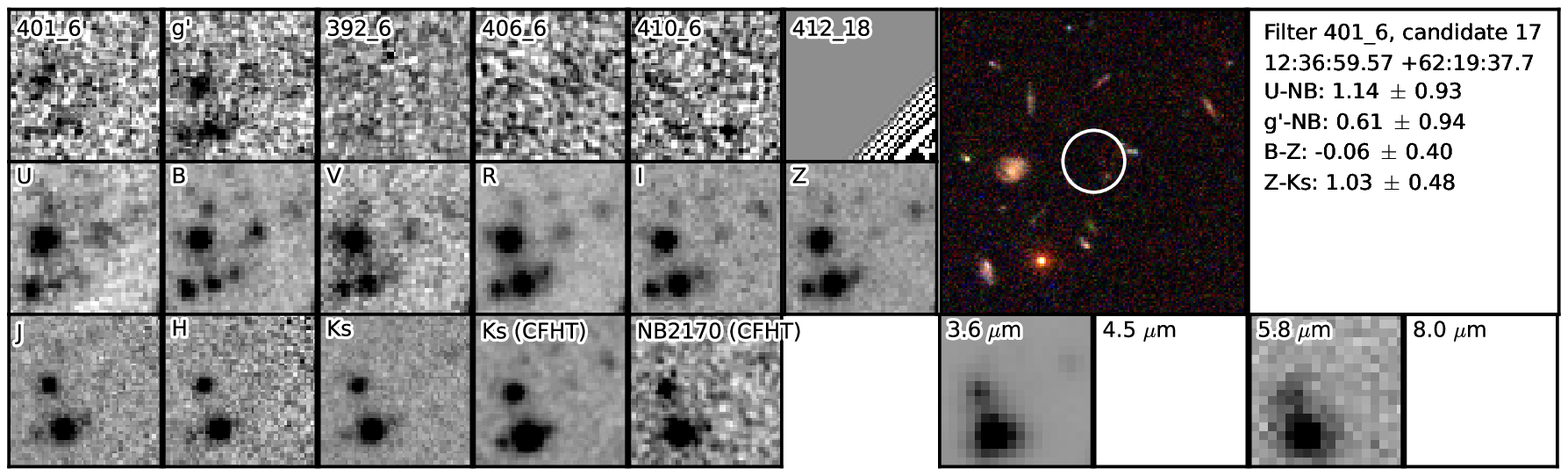}
\includegraphics{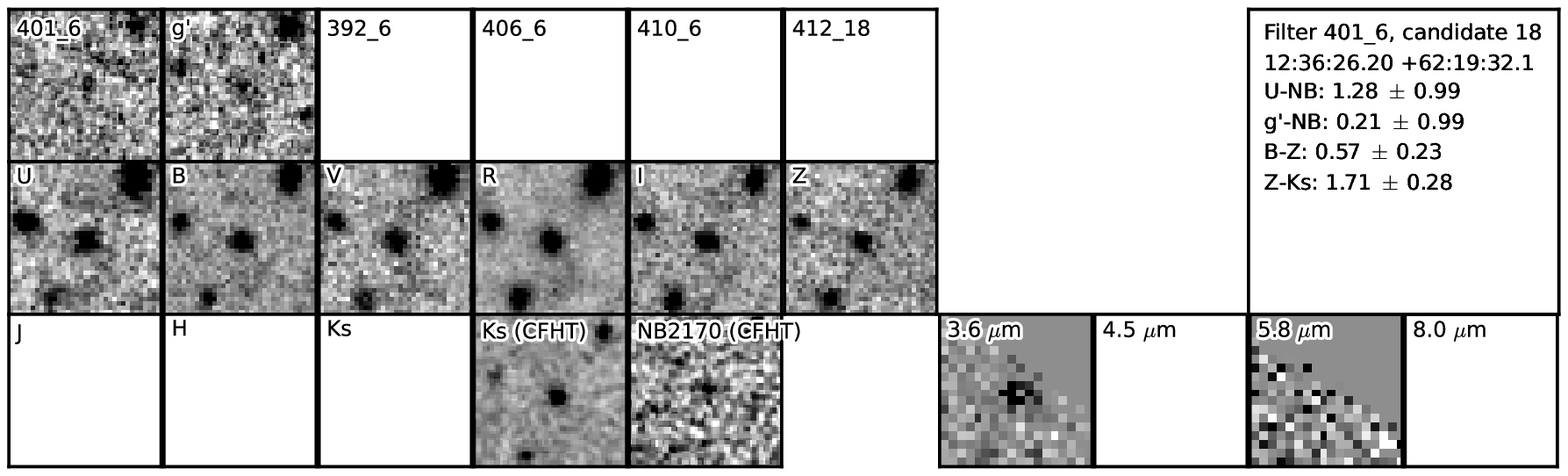}
\includegraphics{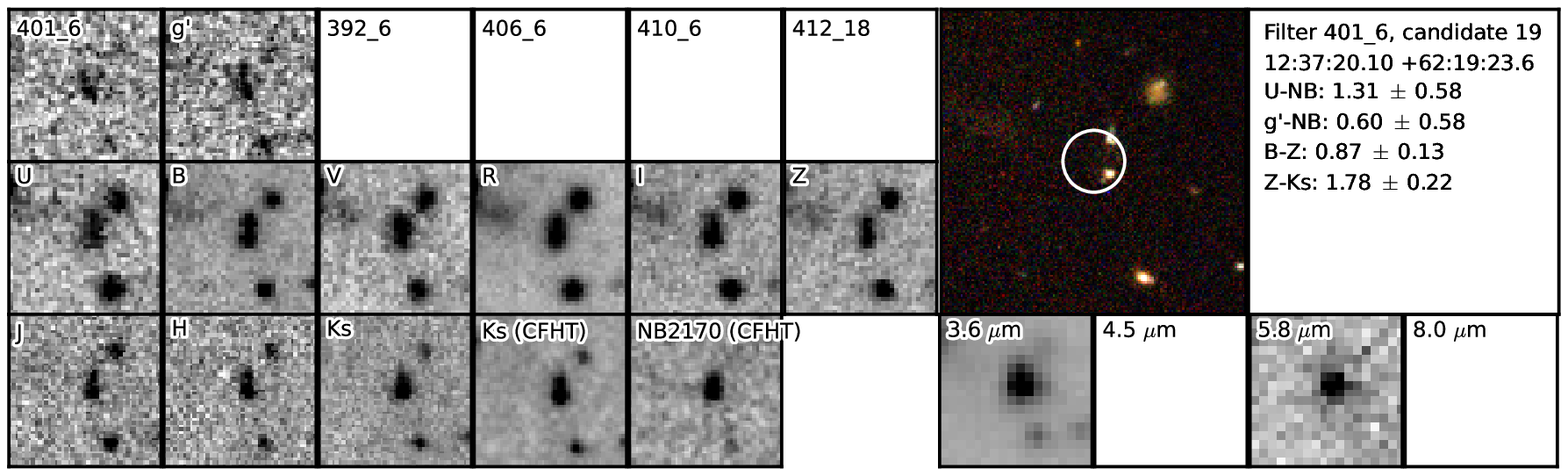}
\includegraphics{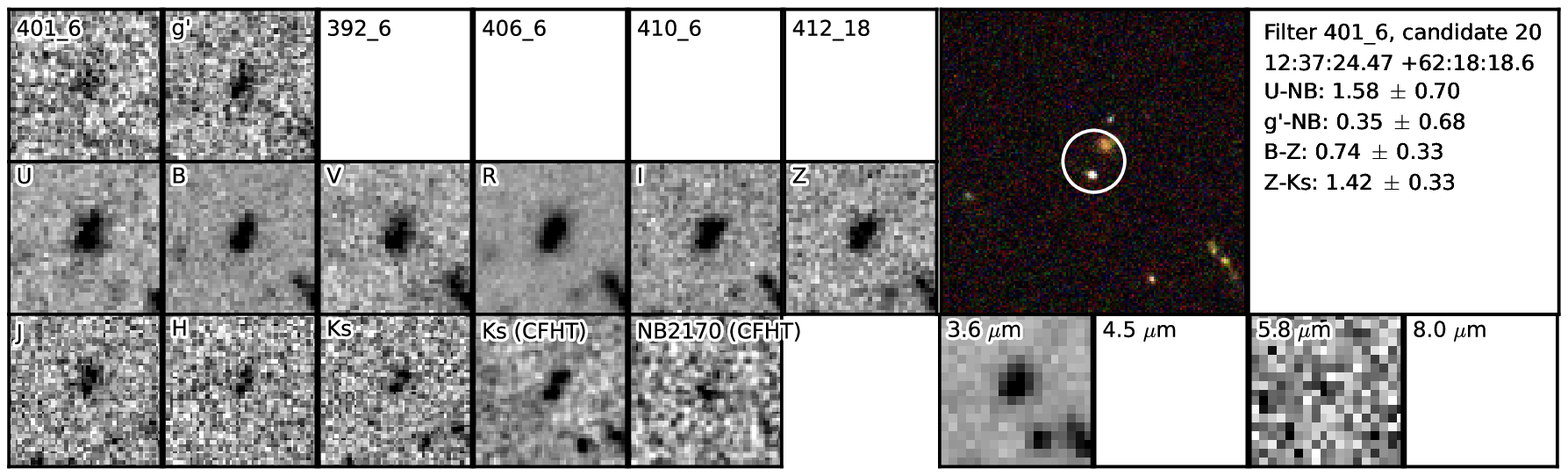}
\caption{Stamps (cont.)}
\end{figure*}

\begin{figure*}
\centering
\includegraphics{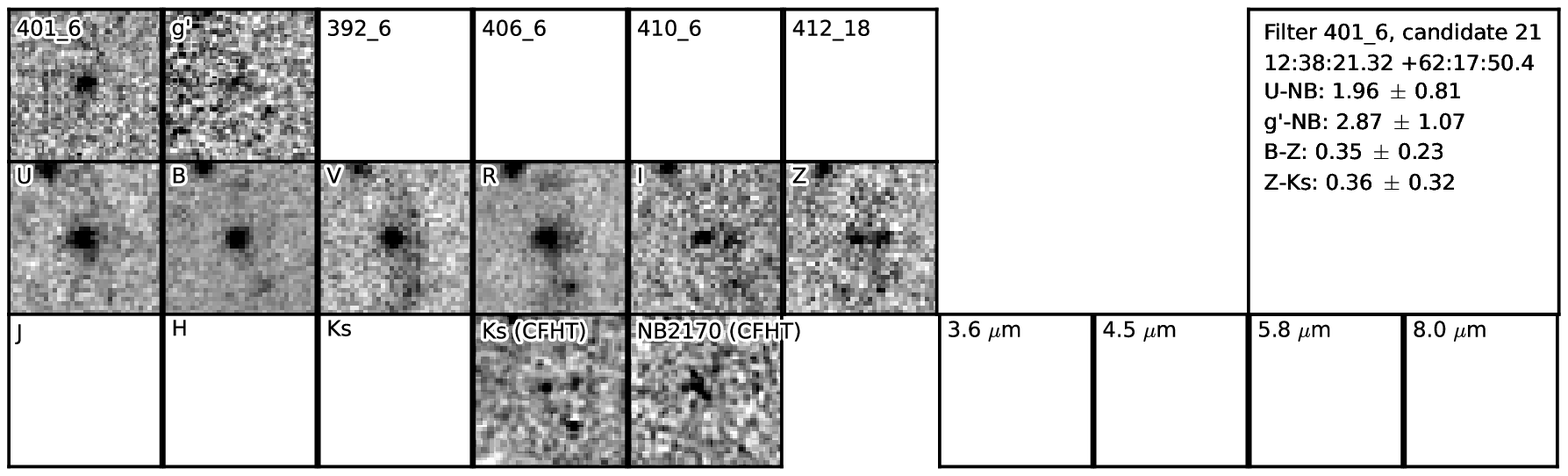}
\includegraphics{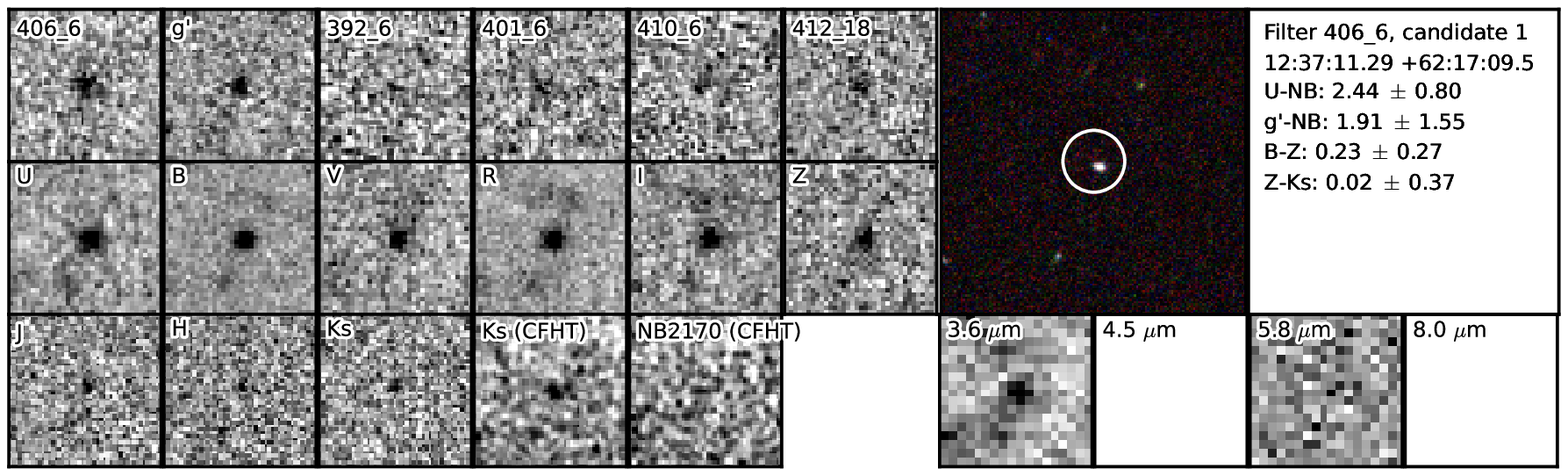}
\includegraphics{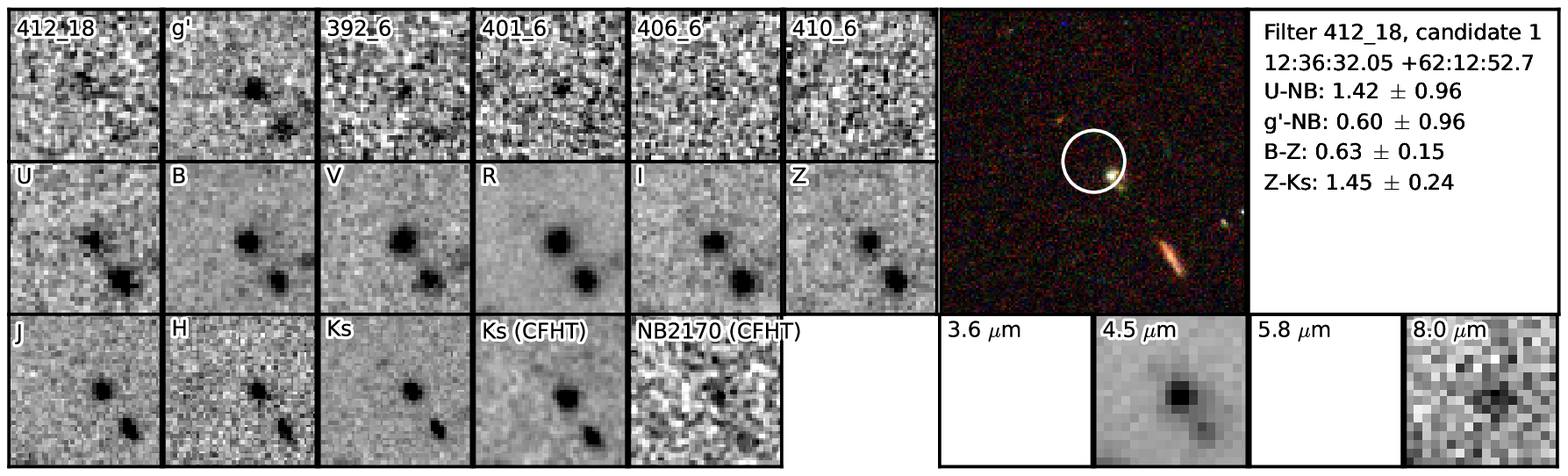}
\caption{Stamps (cont.)}
\end{figure*}

\end{appendix}

\end{document}